\documentclass[aps,prd,twocolumn,10pt,superscriptaddress,nofootinbib,nobibnotes,longbibliography,floatfix]{revtex4-1}

\usepackage{amssymb}
\usepackage{graphicx}
\usepackage{amsmath}
\usepackage{hyperref}
\usepackage{subfigure}
\usepackage{multirow}
\usepackage{setspace}
\usepackage{verbatim}
\usepackage{color}
\usepackage{ulem}
\usepackage[utf8]{inputenc}
\usepackage[table,xcdraw]{xcolor}
\usepackage{makecell}
\allowdisplaybreaks[4]

\newcommand{\md}{\mathrm{d}}
\newcommand{\me}{\mathrm{e}}

\begin{document}

\title{Primordial black hole formation in bulk-viscous cosmology}

\author{Zi-Yan Yuwen}
\email{ziyan.yuwen@apctp.org}
\affiliation{Asia Pacific Center for Theoretical Physics (APCTP), Pohang 37673, Korea}

\author{Cristian Joana}
\email{cristian.joana@ucas.ac.cn}
\affiliation{International Centre for Theoretical Physics Asia-Pacific (ICTP-AP), University of Chinese Academy of Sciences, 100190 Beijing, China}
\affiliation{Center for Theoretical Physics, School of Physics and Optoelectronic Engineering, Hainan University, Haikou 570228, China}

\author{Shao-Jiang Wang}
\email{schwang@itp.ac.cn}
\affiliation{Institute of Theoretical Physics, Chinese Academy of Sciences, Beijing 100190, China}
\affiliation{Asia Pacific Center for Theoretical Physics (APCTP), Pohang 37673, Korea}

\author{Rong-Gen Cai}
\email{caironggen@nbu.edu.cn}
\affiliation{Institute of Fundamental Physics and Quantum Technology \& School of Physical Science and Technology, Ningbo University, Ningbo 315211, China}

\begin{abstract}
We investigate primordial black hole (PBH) formation in a cosmological background with bulk viscosity. Using numerical simulations, we determine the collapse threshold and the resulting PBH mass. We find that the critical threshold $\mu_c$ retains a dependence on the equation-of-state parameter $w$ similar to that in the inviscid case, but is enhanced by an amount comparable to the bulk-viscosity strength $\epsilon$. For fixed $w$, the increase in $\mu_c$ is approximately linear in $\epsilon$. By fitting the standard critical-scaling law for near-threshold collapse, we find that the bulk viscosity leads to an enhancement in the resulting PBH mass. These results indicate that bulk viscosity can systematically modify both the PBH threshold and PBH mass scaling law in the early Universe.
\end{abstract}
\maketitle

\textit{\textbf{Introduction.}---} Primordial black holes (PBHs)~\cite{Zeldovich:1967lct,Hawking:1971ei,Carr:1974nx,Carr:1975qj} are one of the most popular candidates serving as the dark matter~\cite{Carr:2009jm,Carr:2016drx,Carr:2020gox,Carr:2020xqk,Green:2020jor,Carr:2021bzv}, which can be naturally generated by many scenarios such as large curvature perturbations either with positive~\cite{Niemeyer:1999ak,Shibata:1999zs,Passaglia:2021jla,Yoo:2021fxs,Germani:2025hcu} or negative~\cite{Joana:2025gqf} amplitude, (p)reheating dynamics~\cite{Green:2000he,Jedamzik:2010dq,Cotner:2018vug,Martin:2019nuw,Kou:2019bbc}, first-order phase transitions (FOPT)~\cite{Hawking:1982ga,Crawford:1982yz,Moss:1994iq,Khlopov:1998nm,Khlopov:1999ys,Jung:2021mku}, and topological defects~\cite{Hawking:1987bn,Polnarev:1988dh,Garriga:1992nm,Widrow:1989vj,Rubin:2000dq}, etc. (see~\cite{LISACosmologyWorkingGroup:2023njw,Escriva:2022duf,Carr:2020xqk,Yoo:2022mzl} for recent reviews), and may have interaction with a variety of cosmological phenomena spanning both the extreme early Universe (FOPT~\cite{Moss:1984zf,Hiscock:1987hn,Jinno:2023vnr,Yuwen:2024gcf,Wang:2025hwc}, stochastic gravitational wave backgrounds~\cite{Bugaev:2010bb,Dong:2015yjs,Ferrante:2022mui,Zeng:2025law,Ning:2025ogq,Ning:2025yvj,Ning:2026nfs}) and the late-time universe observations (epoch of reionization~\cite{Clark:2018ghm,Mena:2019nhm,Zhao:2024jad,Zhao:2025ddy,Yin:2026hvw}, gravitational lensing~\cite{Griest:2013aaa,Niikura:2017zjd,Niikura:2019kqi,Blaineau:2022nhy,Oguri:2022fir,Cai:2022kbp}). In the most widely studied scenario, large perturbations generated by quantum effects were stretched outside the Hubble horizon during inflation
\cite{Carr:1975qj,Garcia-Bellido:1996mdl,Alabidi:2009bk,Kawasaki:2012wr,Clesse:2015wea,Garcia-Bellido:2016dkw,Carr:2017edp,Dimopoulos:2019wew}, followed by reentering the Hubble horizon during radiation-dominated epoch, and then collapsed due to the nonlinear gravitational interactions. 

Within the standard framework, the collapsing cosmic fluid is usually simply modeled as a perfect fluid with a constant equation-of-state (EoS) parameter $w$. While this assumption greatly simplifies the dynamics, the formation of PBH is intrinsically a highly nonlinear gravitational phenomenon, involving rapid compression, large velocity gradients, and apparent horizon formation~\cite{Niemeyer:1997mt,Niemeyer:1999ak,Shibata:1999zs}. In such extreme regimes, dissipative effects are not negligible (if exist) and could qualitatively modify dynamics of cosmic fluid. Among possible dissipative corrections, viscosity provides the leading-order modification in a hydrodynamical system~\cite{Maartens:1995wt,Zimdahl:1996ka,Bemfica:2020zjp}. 
On the other hand, nonlinear curvature perturbations can source effective imperfect-fluid stresses in a coarse-grained description~\cite{Baumann:2010tm,Ballesteros:2011cm,Giovannini:2015uia}, naturally motivating phenomenological studies of viscous effects.
Since a simplified PBH formation proceeds through spherical overdensities undergoing an isotropic compression (see~\cite{Escriva:2024lmm,Escriva:2024aeo} for anisotropic collapsing resulting PBHs with spins), shear viscosity can be neglected, and thereby bulk viscosity constitutes a natural and well-motivated extension beyond the ideal fluid description. In the early Universe, effective bulk viscosity can arise from various mechanisms, including particle productions, interactions between multicomponent fluids, or coarse-grained descriptions of microscopic degrees of freedom~\cite{Singh:2011dw,Eshaghi:2015tqa,Buoninfante:2016ixe,Paul:2025rqe}. On the other hand, introducing bulk viscosity to an FLRW universe modifies the evolution of Hubble parameter~\cite{Avelino:2013mua,Avelino:2013wea} and provides a potential solution to the Hubble tension~\cite{Normann:2016zby,Brevik:2017msy,Normann:2021bjy}, which motivates us to study a viscous system in cosmology.

In this \textit{Letter}, we incorporate bulk viscosity into the numerical relativity framework and investigate its impact on PBH formation using fully nonlinear numerical simulations. Quantitatively, our analysis focuses on how viscous effects change the threshold of critical collapsing, as well as the scaling law of the PBH mass function. We provide an analytical estimation to reproduce the power-law relation between the threshold and the viscosity found in the numerical results.
Throughout this paper, we take the signature of metric as $(-+++)$ and set $c=G=1$.

\begin{figure}
    \centering
    \includegraphics[width=0.95\linewidth]{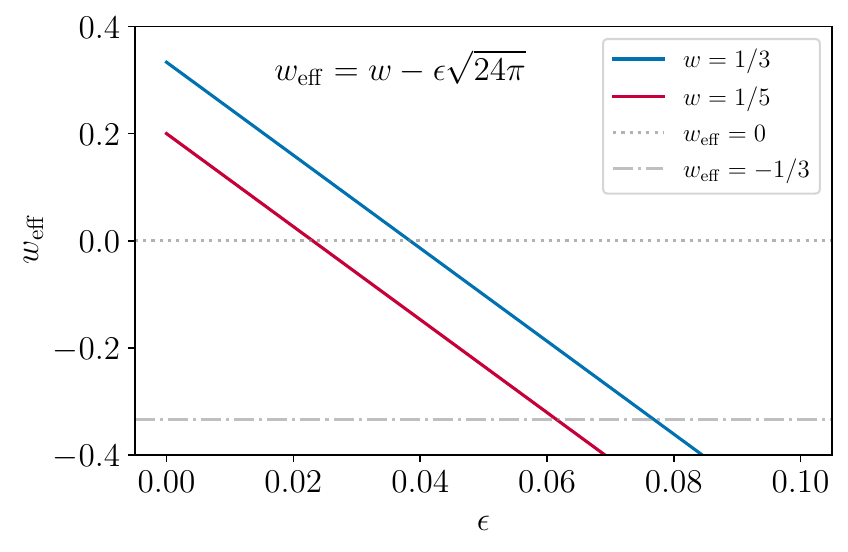}
    \caption{Effective EoS parameter as a function of $\epsilon$ at the homogeneous background level. The intersections of the solid curves with the dotted and dot-dashed horizontal lines give the upper bounds on $\epsilon$ from the near-equilibrium condition and the strong energy condition, respectively.}
    \label{fig: weff}
\end{figure}

\textit{\textbf{Simulations.}---} 
We simulate a spherical symmetric cosmological system using Baumgarte-Shapiro-Shibata-Nakamura (BSSN) formalism~\cite{Baumgarte:1998te,Shibata:1995we}, where the metric take the following $3+1$ decomposition~\cite{Alcubierre:2011pkc},
\begin{align}
    \md s^2 = -\alpha^2 \md t^2 + \me^{4\chi} \left( a \left(\md r - \beta^r \md t \right)^2 + b r^2\md\Omega^2 \right)~.
\end{align}
On the initial time slice, the metric component $\chi$ is associated with a curvature perturbation $\zeta=2\chi$ on superhorizon scales. We consider two initial profiles for $\zeta$: a sinc function~\cite{Yoo:2018kvb,Yoo:2020dkz} and a Gaussian function~\cite{Musco:2018rwt},
\begin{align}
    \zeta^{(s)}(r) = \mu^{(s)} ~\mathrm{sinc}(k_*r)~,~
    \zeta^{(g)}(r) = \mu^{(g)} \me^{-(k_* r)^2}~,
\end{align}
where $k_*$ is the characteristic scale of the perturbation.
Regarding the hydrodynamics, we consider a modification to perfect fluid by the first-order Eckart's viscosity~\cite{Eckart:1940}, in which the stress tensor of the fluid reads
\begin{align}
    T_{\mu\nu} = \left( \rho + p + \Pi\right) u_\mu u_\nu +  \left( p + \Pi\right) g_{\mu\nu}~,
\end{align}
where $\Pi=-\epsilon \rho^{1/2}\Theta_\mathrm{f}$ denotes the bulk viscosity, $\epsilon$ is the dimensionless viscosity strength, and $\Theta_\mathrm{f}=\nabla_\mu u^\mu$ is the expansion scalar of the fluid. 
The detailed numerical implementations can be found in the Supplemental Material~\cite{footnote}. The pressure is related to the energy density by a simple EoS $p=w\rho$. For a homogeneous cosmological background $\Theta_\mathrm{f} = 3H$, the bulk viscosity is then proportional to energy density $\Pi \propto \rho$, effectively behaving as a negative pressure term that softens the fluid according to $w_\mathrm{eff} = w - \epsilon\sqrt{24\pi}$. It should be reminded that the Eckart’s viscosity is no longer expected to provide a reliable description for far-from-equilibrium system with $\Pi\gtrsim p$~\cite{Maartens:1995wt,Maartens:1996dk}, or effectively $p_\mathrm{eff}\gtrsim 0$, which places an upper bound on viscosity strength $\epsilon\lesssim w/\sqrt{24\pi}$. For a radiation-dominated background with $w=1/3$, this gives $\epsilon_{\rm max}\simeq0.038$. A looser upper bound can be obtained by formally extending Eckart's theory into the far-from-equilibrium regime while still imposing the strong energy condition, $\rho+3p_{\rm eff}\geq0$, which yields $\epsilon\leq0.072$. The relation is shown in Fig.~\ref{fig: weff}.

\textit{\textbf{Critical thresholds.}---}
The critical threshold $\mu_c$ for PBH formation depends on the EoS parameter $w$ and viscosity strength $\epsilon$. 
At the background level, the viscosity slows down the cosmic expansion rate, reduces the fluid's effective EoS parameter. However, in the non-perturbative overdense region, the viscosity slows down the contraction rate, increases the dynamical friction which opposes the overdensities from collapsing. 
Therefore, intuitively speaking, the threshold is affected by the competition between these two effects. The numerical simulation results show that the nonlinear friction dominates over the slowing of background expansion, resulting in a higher threshold. 

\begin{table}[t]
    \centering
    \footnotesize
    \begin{ruledtabular}
    \begin{tabular}{ccccc}
        $w$ & \multicolumn{2}{c}{sinc} & \multicolumn{2}{c}{Gaussian} \\
        & $\epsilon=0$ & $\epsilon=0.01$ & $\epsilon=0$ & $\epsilon=0.01$ \\
		\hline
        0.05     & 0.18 & 0.19 & 0.24 & 0.25 \\
        0.10     & 0.31 & 0.32 & 0.41 & 0.42 \\
        0.15     & 0.40 & 0.42 & 0.53 & 0.55 \\
        0.20     & 0.48 & 0.50 & 0.63 & 0.65 \\
        0.25     & 0.54 & 0.57 & 0.70 & 0.72 \\
        $1/3$    & 0.60 & 0.64 & 0.80 & 0.83 \\
    \end{tabular}
    \end{ruledtabular}
    \caption{The critical thresholds of PBH formation for sinc and Gaussian initial profiles for various EoS parameters. The table lists the critical thresholds without bulk viscosity $\epsilon=0$, and those obtained with bulk viscosity parameter $\epsilon = 0.01$.}
    \label{tab: critical_thresholds}
\end{table}

\begin{figure}[t]
    \centering
    \includegraphics[width=\columnwidth]{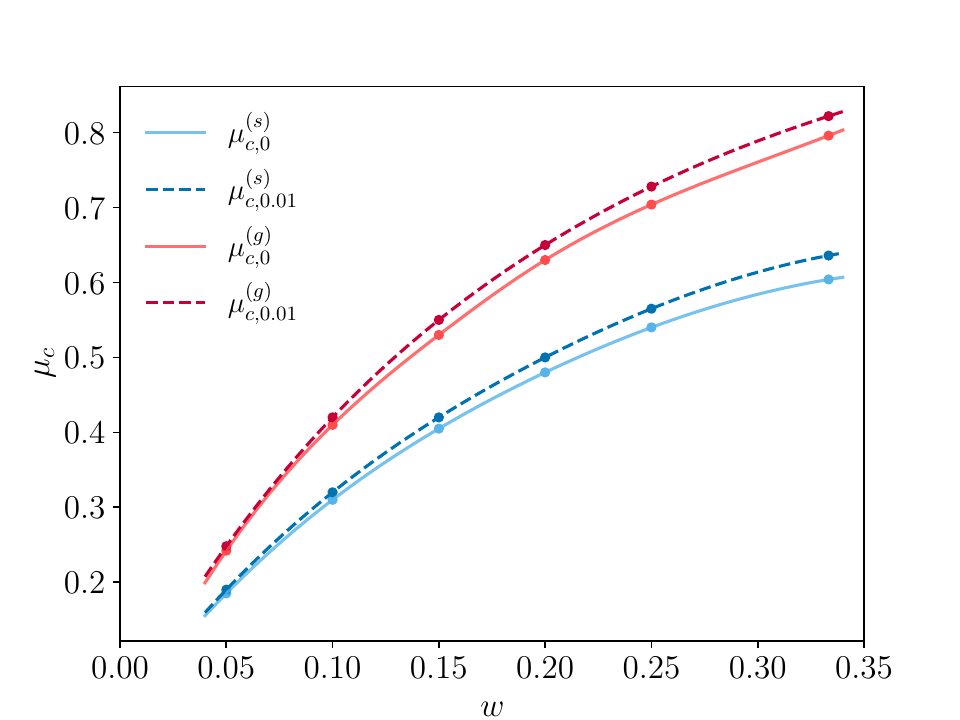}
    \caption{Thresholds $\mu_c$ as functions of EoS parameter, for sinc (blue) and Gaussian (red) initial profiles. Results with bulk viscosity are shown in light solid lines, while those without bulk viscosity are shown in dark dashed lines. Dots represent numerical data points, and the curves are obtained using cubic interpolation.}
    \label{fig: threshold}
\end{figure}

In Table~\ref{tab: critical_thresholds}, we list out the numerical values of $\mu_c$ for different initial profiles with various $w$ and $\epsilon$, which are consistent with the values obtained from full BSSN simulations~\cite{Joana:2025gqf} and standard Misner-Sharp simulations~\cite{Ning:2025yvj}. We plot the data points and present the $w$-dependence of $\mu_c$ by a cubic interpolation in Fig.~\ref{fig: threshold}, showing a clear monotonically increasing threshold. It can also be seen from this figure that the threshold for the sinc initial profile is always smaller than that for the Gaussian initial profile, which can be attributed to the fact that the sinc initial profile stores more tension in the gradient of $\zeta$.

\begin{figure*}[t]
    \centering
    \includegraphics[width=0.45\textwidth]{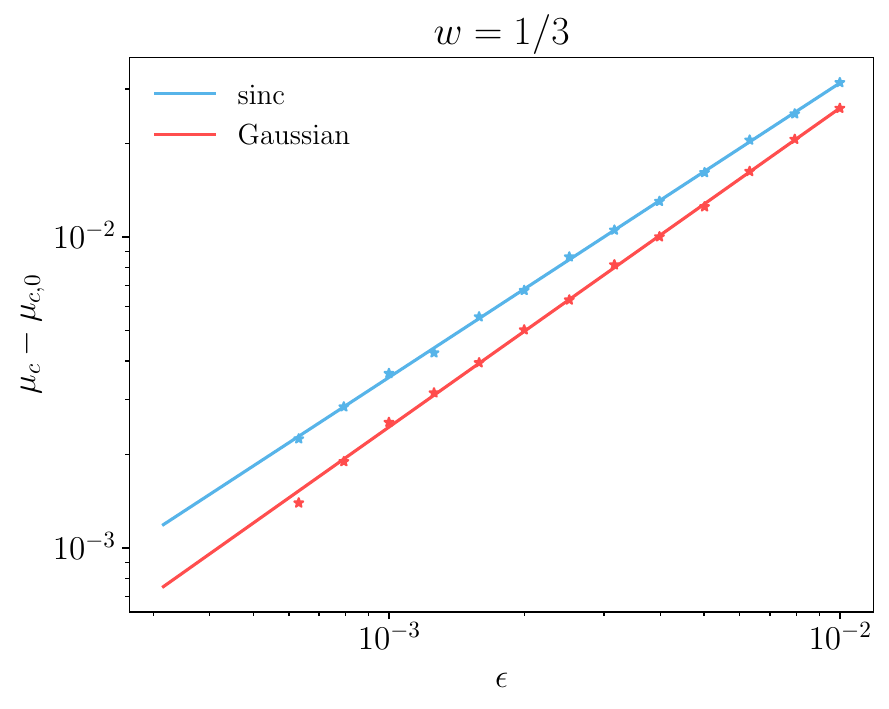}\quad
    \includegraphics[width=0.45\textwidth]{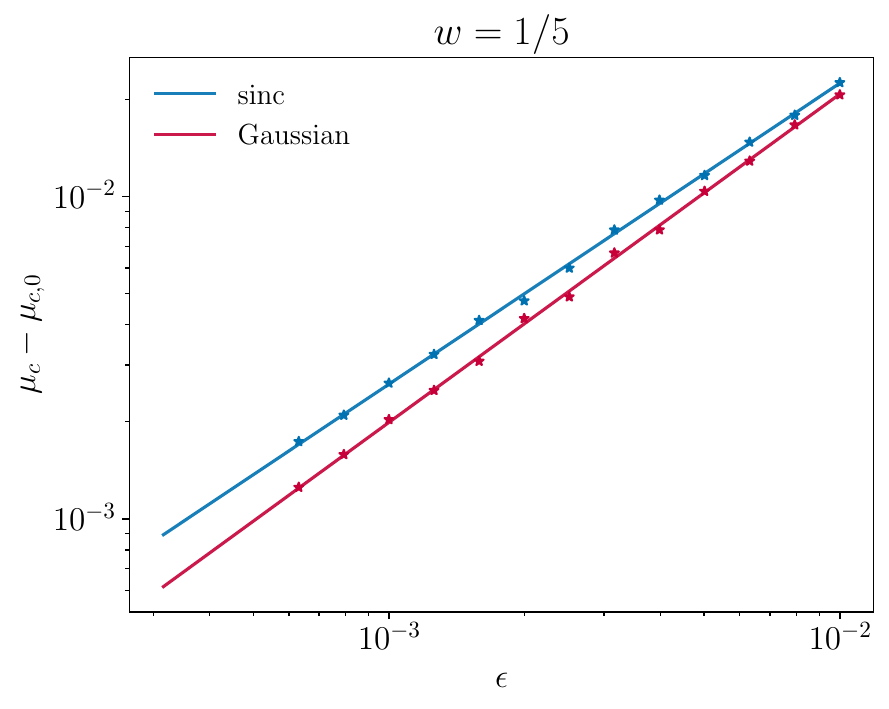}\quad
    \caption{The power law of critical threshold functions of viscosity strength $\epsilon$, for sinc (blue) and Gaussian (red) initial profiles, where $\mu_{c,0}$ denotes the threshold with $\epsilon=0$. The results with $w=1/3$ are shown in the left panel, while those with $w=1/5$ are shown in the right panel.}
    \label{fig: threshold power law}
\end{figure*}

It is noticed that the threshold is enhanced by a value of the same order of magnitude as $\epsilon$. This enhancement is found to be well fitted by a power law as follows,
\begin{align}
    \mu_c - \mu_{c,0} = K \epsilon^\Gamma~,
\end{align}
where $\mu_{c,0}$ is the threshold without viscosity, $K$ and $\Gamma$ are the amplitude and power index, respectively. In Fig.~\ref{fig: threshold power law}, we present the power-law fittings for two initial profiles with two typical EoS parameters $w=1/3$ and $1/5$, respectively. The best fittings are provided in Table~\ref{tab: threshold power law}. 
In all cases, it is found that the scaling index $\Gamma$ is very close to $1$, and the amplitude is $ K \sim \mathcal{O}(1)$, confirming that the enhancement $\mu_c - \mu_{c,0}$ is of the same order as $\epsilon$. 
For $w=1/3$ and $\epsilon=0.01$, the background EoS parameter is reduced to $w_{\rm eff}\simeq 0.247$, approximately $3/4$ of its original value. Thus, the near-linear scaling remains accurate over the physically relevant viscosity range, where the viscous correction to the pressure is already not negligible, rather than being restricted to the limit of infinitesimal viscosity.

\begin{table}[t]
    \centering
    \footnotesize
    \begin{ruledtabular}
    \begin{tabular}{ccccc}
        $w$ & \multicolumn{2}{c}{sinc} & \multicolumn{2}{c}{Gaussian} \\ 
        & $K$ & $\Gamma$ & $K$ & $\Gamma$ \\
        \hline
        $1/3$ & 2.48 & 0.95 & 2.94 & 1.03 \\
        $1/5$ & 1.68 & 0.94 & 2.27 & 1.02 \\
    \end{tabular}
    \end{ruledtabular}
    \caption{Best-fit parameters for the power law of the threshold enhancement shown in Fig.~\ref{fig: threshold power law}}
    \label{tab: threshold power law}
\end{table}

\textit{\textbf{Scaling law of mass functions.}---}
The mass of the black hole $M_\mathrm{PBH}$ near critical collapsing follows a  standard scaling law of the following form~\cite{Choptuik:1992jv,Evans:1994pj,Musco:2012au}
\begin{align}\label{eq: mass function}
    \frac{M_\mathrm{PBH}}{M_\times} = k|\mu - \mu_c|^\gamma~,
\end{align}
where $M_\times$ is the mass within the Hubble volume at the time of horizon-crossing, $\mu_c$ is the critical threshold. Both $M_\mathrm{PBH}$ and $M_\times$ are obtained by computing the Misner-Sharp-Hernandez mass, the former of which corresponds to an areal radius of the outermost trapping surface at the time of apparent horizon formation, while the latter of which corresponds to an areal radius $R=\me^{2\chi}\sqrt{b} \>r$ satisfying $k_* r=1$ at the time of horizon-crossing $R=1/H$. $k$ and $\gamma$ are the amplitude and power index, respectively, which can be obtained by a power law fitting. 

\begin{figure*}[t]
    \centering
    \includegraphics[width=0.45\textwidth]{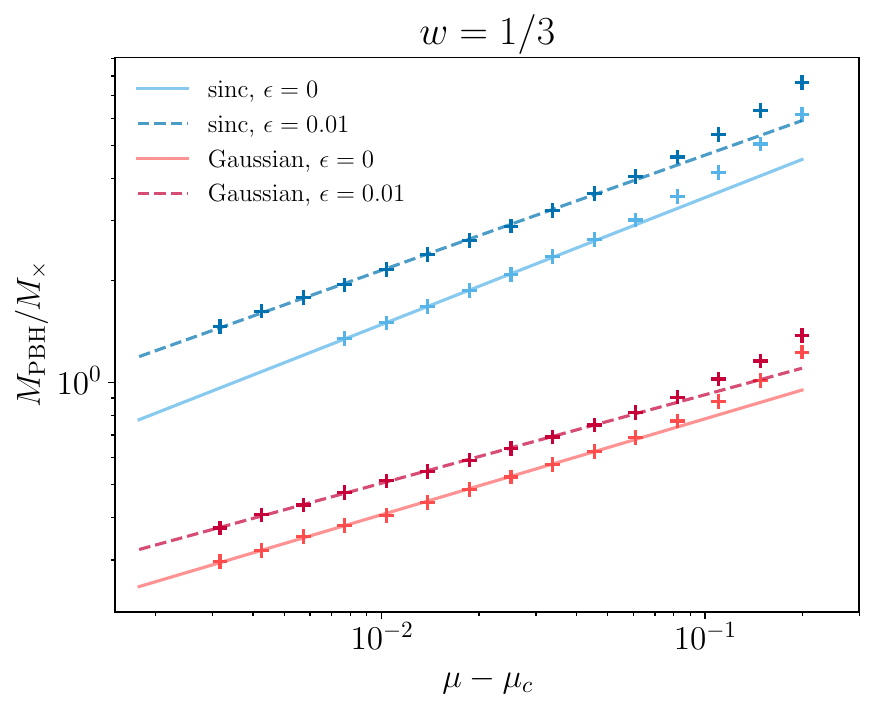}\quad
    \includegraphics[width=0.45\textwidth]{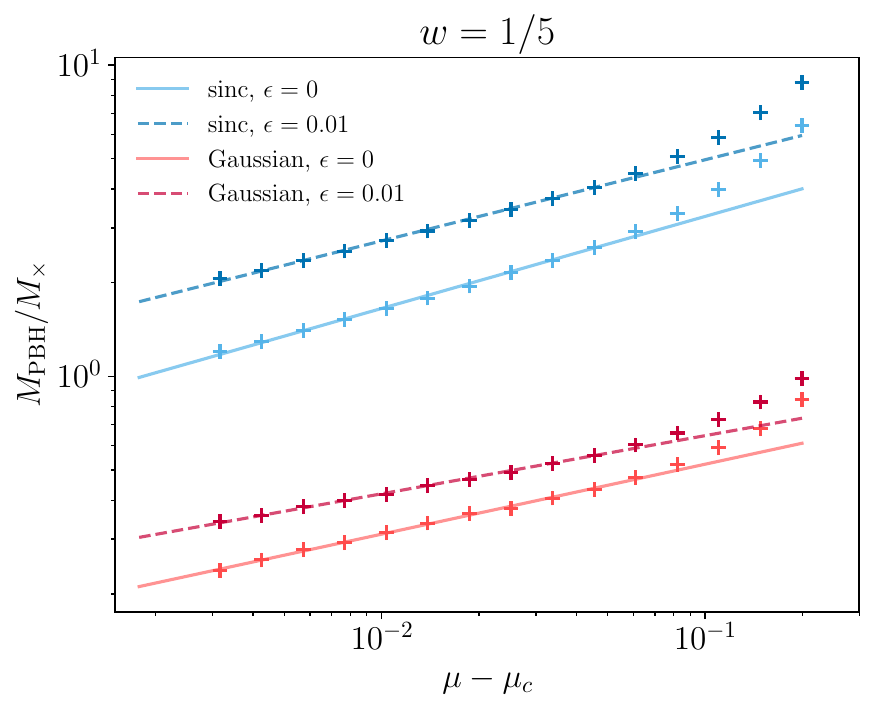}\quad
    \caption{The power law of PBH mass formation with and without bulk viscosity, for $w=1/3$ (left) and $w=1/5$ (right), respectively. The results for sinc initial profiles are shown in blue, while those for Gaussian initial profiles are shown in red.}
    \label{fig: scaling law}
\end{figure*}

The best fitted parameters are provided in Table~\ref{tab: scaling law}, and the corresponding data point and fitted scaling law are presented in Fig.~\ref{fig: scaling law}. It should be noted that the scaling law is valid only in the vicinity of critical collapsing, that is, $\mu - \mu_c \ll 1$. In the case of PBH formation with sufficiently strong perturbations ($\mu - \mu_c \gtrsim 0.1$)~\cite{Escriva:2022duf,Joana:2025gqf}, the resulting black hole mass can exceed the value predicted by the scaling law by approximately a factor of two. Although the $\gamma$ parameter remains at a similar order of magnitude, comparing the best fitted value of $k$ for different initial profiles, it is found that the sinc initial profiles always result in a higher mass ratio than the Gaussian initial profile by around one order of magnitude. This is expected because the density contrast decays exponentially beyond the length scale for the Gaussian case, while it falls off as $1/r$ for the sinc initial profiles.

\begin{table}[t]
    \centering
    \footnotesize
    \begin{ruledtabular}
    \begin{tabular}{cccccc}
        $w$ & $\epsilon$ & \multicolumn{2}{c}{sinc} & \multicolumn{2}{c}{Gaussian} \\
        & & $k$ & $\gamma$ & $k$ & $\gamma$ \\
        \hline
        $1/3$ & 0    & 8.30 & 0.37 & 1.50 & 0.28 \\
        $1/3$ & 0.01 & 10.22 & 0.34 & 1.68 & 0.26 \\
        $1/5$ & 0    & 5.52 & 0.26 & 0.88 & 0.23 \\
        $1/5$ & 0.01 & 9.05 & 0.26 & 0.99 & 0.19 \\
    \end{tabular}
    \end{ruledtabular}
    \caption{Best-fit parameters for the power-law mass function of critical collapsing shown in Fig.~\ref{fig: scaling law}}
    \label{tab: scaling law}
\end{table}

\textit{\textbf{Analytical estimation.}---}
The nearly linear scaling law of PBH threshold in the bulk viscosity parameter can be understood analytically with the following estimations. Consider a cosmic fluid with $p=w\rho$, the viscosity dilutes the pressure as $p_\mathrm{eff}=p-\xi\Theta_\mathrm{f}$. In an expanding background, the expansion parameter looks like $\Theta_\mathrm{f}=\nabla_\mu u^\mu\simeq 3H+\nabla\cdot v$. Near horizon crossing, $H\simeq t_H^{-1}$, we can define a dimensionless viscosity parameter $\delta w=\xi H/\rho = \epsilon\sqrt{8\pi /3}$. This allows the effective pressure to be rewritten as $p_\mathrm{eff}=w\rho-3\xi H=w\rho(1-3\delta w/w)$, corresponding to a softened fluid's EoS $w_\mathrm{eff}=p_\mathrm{eff}/\rho=w-3\delta w$. 

Although viscosity softens the effective EoS, it also introduces dissipative damping that tends to stabilize against the collapse, which made the dominant contribution to the PBH formation threshold shift. The main difference is that, in the overdense region, the fluid tends to collapse instead of expansion, i.e. $\Theta_\mathrm{f}<0$, thereby leads to an increase in the effective pressure. The continuity equation indicates that the growth rate of the overdensity is approximately given by the expansion of the fluid $D_\tau{\delta}\simeq -(1+w)(1+\delta)\Theta_\mathrm{f}$, where $\delta=\delta\rho/\bar{\rho}$ is the local density contrast, and $D_\tau$ denotes for the derivative with respect to the fluid proper time. The characteristic compression time scale can be estimated through the growth rate of the density contrast
\begin{align}
    t_{\mathrm{comp}}^{-1} = D_\tau\ln(1+\delta) \simeq -(1+w)\Theta_\mathrm{f}~.
\end{align}
Note that the change to the pressure due to viscosity is $\Delta p_\mathrm{vis} = -\epsilon \rho^{1/2} \Theta_\mathrm{f}$, the effective sound speed is then modified as
\begin{align}
    c_s^2 = c_{s,0}^2 + \frac{\Delta p_\mathrm{vis}}{\rho}\simeq w + \frac{\epsilon}{\rho^{1/2}(1+w) t_{\mathrm{comp}}}~.
\end{align}
On the other hand, the collapsing time scale is given by an approximate balancing between the gravity compression and the decompression caused by the pressure. At background level, the inverse of compression time is approximately equal to the sound crossing time, $t_{\mathrm{comp}}^{(0)}\simeq t_{\mathrm{sound}}^{(0)} \simeq R_*/c_s^{(0)}= R_*/\sqrt{w}$, where $R_*$ is the characteristic scale of the perturbation and can be replaced by the Hubble horizon at entering time $H_*$ for estimation. Using the Friedman equation $H_*^2 = 8\pi \rho/3$, to the leading order the effective sound velocity in the overdense region is estimated as
\begin{align}
    c_{s,\mathrm{eff}}^2 &\simeq w + \frac{\epsilon}{\rho^{1/2} (1+w)(t_{\mathrm{comp}}^{(0)} + \mathrm{higher~order})} \nonumber\\
	&\simeq w + \frac{\epsilon}{\rho^{1/2} (1+w)t_{\mathrm{sound}}^{(0)}} \simeq w + \sqrt{\frac{8\pi w}{3}}\frac{\epsilon}{1+w} ~,
\end{align}
which suggests a harder EoS in this region rather than a softer one in the perturbative region. 

The critical collapse corresponds to the case where the gravitational collapsing is balanced by the support of the pressure, $t_\mathrm{grav}(\delta_c) = t_\mathrm{pressure}(\epsilon)\simeq R_*/c_{s,\mathrm{eff}}$. Expanding this balance $t_\mathrm{grav}^{-2}(\delta_c) = t_\mathrm{pressure}^{-2}(\epsilon)$ to the linear order leads to
\begin{align}
    t_\mathrm{grav}^{-2}(\delta_{c,0}) + \frac{\partial t_\mathrm{grav}^{-2}}{\partial \delta_c}\bigg|_{\delta_{c,0}} \Delta\delta_c  = \frac{w}{R_*^2} + \sqrt{\frac{8\pi w}{3}}\frac{\epsilon}{(1+w)R_*^2}~.
\end{align}
Physically, a higher overdensity provides stronger gravitational attraction and indicates shorter collapsing time, i.e. smaller $t_\mathrm{grav}$, thus one has $F \equiv\partial_\delta t_\mathrm{grav}^{-2} >0$. A rough estimation can be stated as follows. 
The gravitation collapsing time can be estimated from the Jeans argument for PBH collapse~\cite{Harada:2013epa}, $t_\mathrm{grav}\sim 1/\sqrt{4\pi \delta\rho}$, yielding
\begin{align}
    F \equiv \frac{\partial t_\mathrm{grav}^{-2}}{\partial \delta} \simeq 4\pi \frac{\partial\delta\rho}{\partial\delta} = 4\pi \bar{\rho} > 0~.
\end{align}

As a result, the threshold change $\Delta\delta_c$ is proportional to $\epsilon$, together with a positive pre-factor,
\begin{align}
    \Delta \delta_c \simeq\sqrt{\frac{8\pi}{3}}\frac{\sqrt{w}}{(1+w)R_*^2 F} \, \epsilon~.
\end{align}
This is consistent with our numerical results: viscosity leads to a higher threshold, and a higher $w$ corresponds to a larger pre-factor. 
Note that $F$ should also exhibit a weak dependence on $w$ and the initial profile of the perturbation, the dependence of $\Delta \delta_c$ on $w$ is thereby not simply proportional to $\sqrt{w}/(1+w)$, although it should remain approximately positively correlated with $w$. 
At horizon crossing time where the scale of perturbation is comparable to the Hubble horizon, the density contrast and the amplitude of curvature perturbation are related by a simple proportion relation $\delta = \frac{2(1+w)}{5+3w}\mu$. Therefore, a linear dependence of $\Delta\delta_c$ on $\epsilon$ directly implies a linear scaling $\Delta \mu_c \propto \epsilon$.
Finally, the same reasoning suggests that the PBH mass scaling relation should acquire a shift in the critical amplitude,
\begin{align}
    M_{\mathrm{PBH}}=M_H\,
    \left(
        \delta - \delta_{c,0}(1+K \epsilon)
    \right)^{\gamma},
\end{align}
where the dependence of the horizon mass $M_H$ on $\epsilon$ encodes the modification of the background dynamics induced by the viscous fluid. 

\textit{\textbf{Conclusions and discussions.}---}
In this \textit{Letter}, we systematically investigate PBH formation in a viscous universe using numerical simulations. In particular, we focus on near-critical collapse to obtain the relation between the PBH formation threshold $\mu_c$ and the fluid intrinsic properties: the EoS parameter $w$ and the bulk-viscosity strength $\epsilon$. 
The viscosity increases the threshold for PBH formation by dissipation, meanwhile keeping the similar dependence behavior on EoS. By numerical fitting, it is found that the threshold enhancement follows a near-linear power law dependence on $\epsilon$. 
The PBH mass function of critical collapsing can still be described by the standard scaling law, which is also fitted for several typical parameter sets. By introducing viscosity, it is found that the ratio $M_\mathrm{PBH} / M_\times$ is enhanced by an order one factor. A more delicated analytical derivation of the modified scaling law is left for future work.

Although the resulting thresholds enhancements are numerically small, a seemingly modest change in the collapse threshold can lead to a much larger modification of the PBH abundance. Assuming Gaussian perturbations, the PBH formation fraction can be estimated by the rare tail of the distribution $\beta(\mu_c)\propto \mathrm{erfc}\left(-\mu_c/(\sqrt{2}\sigma)\right)$, where $\sigma^2$ is the variance of the amplitude of perturbations. Taking $\sigma^2\simeq 0.01$ (see~\cite{Green:2020jor} for example) as a representative value for PBH formation, a change of the collapse threshold from $\mu_c=0.60$ (inviscid) to $\mu_c=0.64$ (viscous) for radiation-dominated background $w=1/3$, corresponding to only about a $7\%$ shift in $\mu_c$, changes the abundance by $\beta(0.60)/\beta(0.64)\simeq 12$. Therefore, even a small viscosity-induced shift in the collapse threshold can translate into an order-of-magnitude change in the predicted PBH abundance.

By adopting the comoving gauge, we simplify the procedure of adding viscosity into the system. However, the comoving gauge is not numerically stable for large EoS parameters and negative perturbations, and is not suitable for the type-II PBH formation process. Future work can be carried out by considering simulations with viscosity in a better gauge to improve numerical stability, or evolving the viscosity with a second-order Israel–Stewart theory. Both these extensions would require a more sophisticated time-integration method to properly handle the stiffness in the EoM, which may potentially slow down the simulation. Establishing a consistent numerical-relativity framework for far-from-equilibrium hydrodynamics with strong viscosity beyond the validity of Eckart's theory in a cosmological background remains an open problem.

Other possible future work could involve performing a full $3+1$ simulation that consistently incorporates both viscosity and anisotropy, in order to study how viscous effects influence the spin and ellipticity of the resulting black hole, thereby providing a better understanding of the role of viscosity in the nonlinear dynamics of gravitational collapse. In this case, both bulk and shear viscosity can be introduced to the system for a more complete analysis.

\textit{\textbf{Acknowledgments.}---}
The authors thank Misao Sasaki, Jing Liu, Hao-Tian Sun, Zhuan Ning, and Xiang-Xi Zeng for helpful discussions. 
R.-G. C. and S.-J. W. are supported by the National Key Research and Development Program of China Grants No. 2021YFA0718304, No. 2021YFC2203004, and No. 2020YFC2201501, the National Natural Science Foundation of China Grants No. 12422502, No. 12547110, No. 12588101, No. 12235019, and No. 12447101, and the Science Research Grants from the China Manned Space Project with No. CMS-CSST-2025-A01.
Z.-Y. Yuwen is supported by an appointment to the Young Scientist Training (YST) program at the APCTP through the Science and Technology Promotion Fund and Lottery Fund of the Korean Government. This was also supported by the Korean Local Governments-Gyeongsangbuk-do Province and Pohang City.
C.J. is supported by NSFC grants No. W2433007, E414660101 and No. 12547104, and from the Fundamental Research Funds for the Central Universities under grants No. E4EQ6604X2 and No. E3ER6601A2.



\clearpage
\onecolumngrid
\newpage

\begin{center}
{\Large\textbf{Supplemental Material for\\ Primordial black hole formation in bulk-viscous cosmology}}
\end{center}
\vspace{0cm}

\begin{center}
{\large Zi-Yan Yuwen$^{1,}$\footnotemark[1], 
Cristian Joana$^{2,}$\footnotemark[2], 
Shao-Jiang Wang$^{3,1,}$\footnotemark[3], 
Rong-Gen Cai$^{4,}$\footnotemark[4],}\\
$^{1}${\it Asia Pacific Center for Theoretical Physics (APCTP), Pohang 37673, Korea}\\
$^{2}${\it International Centre for Theoretical Physics Asia-Pacific (ICTP-AP), \\
University of Chinese Academy of Sciences, 100190 Beijing, China}\\
$^{3}${\it Institute of Theoretical Physics, Chinese Academy of Sciences, Beijing 100190, China}\\
$^{4}${\it Institute of Fundamental Physics and Quantum Technology \& School of \\
Physical Science and Technology, Ningbo University, Ningbo 315211, China}\\
\end{center}
\footnotetext[1]{ziyan.yuwen@apctp.org}
\footnotetext[2]{cristian.joana@ucas.ac.cn}
\footnotetext[3]{schwang@itp.ac.cn (corresponding author)}
\footnotetext[4]{caironggen@nbu.edu.cn}

\setcounter{section}{0}
\setcounter{equation}{0}
\setcounter{figure}{0}
\setcounter{table}{0}
\renewcommand{\theequation}{S\arabic{equation}}
\renewcommand{\thefigure}{S\arabic{figure}}
\renewcommand{\thetable}{S\arabic{table}}
\renewcommand{\theHequation}{S\arabic{equation}}
\renewcommand{\theHfigure}{S\arabic{figure}}
\renewcommand{\theHtable}{S\arabic{table}}

\section{Simulation framework}\label{sec: setup}

In this section, we present the physical and mathematical framework adopted in our simulations, including the description for evolving equations, apparent horizon finders, initial data generations, and numerical implementations.

\subsection{Equation of motions}

A classical general relativistic system is characterized by the spacetime metric and the matter stress tensor, with the former governed by Einstein’s equations and the latter constrained by covariant conservation. For the spacetime sector, let us focus on a system with spherical symmetry, which sufficiently reduces the number of independent components of the metric and the Baumgarte-Shapiro-Shibata-Nakamura (BSSN) formalism~\cite{Baumgarte:1998te,Shibata:1995we}, the line-element is given by~\cite{Alcubierre:2011pkc}
\begin{align}\label{eq: metric}
    \md s^2 = -\alpha^2 \md t^2 + \me^{4\chi} \left( a \left(\md r - \beta^r \md t \right)^2 + b r^2\md\Omega^2 \right)
\end{align}
where $\alpha$ and $\beta^i$ are the lapse and shift functions, respectively, and $\md\Omega^2 = \md\theta^2 + \sin^2\theta \md\phi^2$ is the standard metric on a $2$-dim sphere $\md\Omega^2 = \md\theta^2 + \sin^2\theta \md\phi^2$. Under the BSSN formalism, the spatial metric $\gamma_{ij}$ is decomposed into its conformal factor $\chi$ and another spatial metric $\tilde{\gamma}_{ij}$ with a unitary determinant in a Cartesian coordinate. In the spherical coordinate, the non-vanishing components of $\tilde{\gamma}_{ij}$ are given by $\tilde{\gamma}_{rr} = a$ and $\tilde{\gamma}_{\theta\theta} = b r^2$. The extrinsic curvature $K_{ij}$ encodes the ``time derivative'' information of the spatial metric, is decomposed to its trace $K$ and a trace-free tensor $A_{ij}$ as $K_{ij} = \frac{1}{3}K\gamma_{ij} + A_{ij}$. After a similar conformal transformation to the metric, we further perform a conformal rescaling of the traceless extrinsic curvature as $\tilde{A}_{ij} = \exp(-4\chi) A_{ij}$, with two independent components defined as
$A_a = \tilde{A}_r^{r}$, $ A_b = \tilde{A}_\theta^{\theta}$, satisfying the traceless condition $A_a + 2A_b = 0$. In addition, the BSSN formalism also introduces $\hat{\Delta}^i =-\partial_j \tilde{\gamma}^{ji}$ as extra independent evolution variables to maintain numerical stability during simulations~\cite{Baumgarte:1998te}.

For the matter sector, the stress tensor is described by an imperfect fluid with bulk viscosity, whose stress tensor takes the form
\begin{align}
    T_{\mu\nu} = \left( \rho + p + \Pi\right) u_\mu u_\nu +  \left( p + \Pi\right) g_{\mu\nu}~,
\end{align}
where $\rho$, $p$ are the energy density and pressure, respectively, measured by the comoving observer, where the pressure is proportional to the energy density $p=w\rho$ with a positive constant $w$.
$\Pi$ denotes the bulk viscous contribution. In principle, one can also add the first-order shear viscosity to the system. Nevertheless, since the system considered here is spherical symmetric, the shear viscosity does not contribute to the dynamical evolutions and therefore is neglected. The simplest description corresponds to the first-order Eckart's theory~\cite{Eckart:1940}, 
\begin{align}\label{eq: Eckart viscosity}
    \Pi = -\xi \Theta_\mathrm{f}~, \quad \xi\equiv\epsilon \rho^q
\end{align}
where a constant $\epsilon$ measures the strength of viscosity and $\Theta_\mathrm{f} =\nabla_\mu u^\mu $ is the expansion of the fluid. Here we choose the power index to be $q=1/2$, which is a typical value discussed in previous literature~\cite{DiPrisco:2000dw,Carlevaro:2005xw}, and is also a marginal value satisfying the stability condition when considering a viscous inflation~\cite{Maartens:1995wt}. For a cosmological background with $\Theta_\mathrm{f} = 3H$, the bulk viscosity then takes the form $\Pi \propto \rho$, effectively behaving as a negative pressure term. Therefore, at the background level, the bulk viscous fluid admits an effective perfect-fluid description with a shifted EoS parameter $w_\mathrm{eff} = w - \epsilon \sqrt{24\pi}$. A more complete and stable formulation is provided by the Israel–Stewart theory~\cite{Israel:1976tn,Israel:1979wp}. For simplicity, we adopt the bulk viscosity in the form given in Eq.~\eqref{eq: Eckart viscosity} in this work. A detailed discussion on the evolution of bulk viscosity under an arbitrary gauge is provided in Section.~\ref{app: IS}.

In general, the BSSN evolution variables for the fluid sector are defined by the following covariant variables,
\begin{align}
    \mathcal{D} & \equiv \rho_0 W~, \\
    \mathcal{E} & \equiv (\rho + p+\Pi)W^2 -(p+\Pi) - D~, \\
    S_i & \equiv (\rho + p+\Pi)W^2 v_i~,
\end{align}
where $\rho_0$ is the fluid's inertial energy, $v^r$ is the $3$-velocity of the fluid relative to the Eulerian observers, and $W\equiv1/\sqrt{1-\gamma_{ij}v^iv^j}$ is the Lorentz factor. The projection of the stress tensor can then be expressed as
\begin{align}
    \rho_E & = n^\mu n^\nu T_{\mu\nu} = \mathcal{D} + \mathcal{E}~,\\
    S_i & = -\gamma_i^\mu n^\nu T_{\mu\nu}~, \\
    S_{ij} & = \gamma_i^\mu \gamma_j^\nu T_{\mu\nu}~.
\end{align}
The Equation of Motions (EoMs) for the covariant variable set $\{\mathcal{D},\mathcal{E}, S_i\}$ read~\cite{Alcubierre:2008book}
\begin{align}
    \left( \partial_t - \mathcal{L}_\beta \right) \mathcal{D}  & = - D_k (\alpha \mathcal{D} v^k) + \alpha K \mathcal{D}  ~,\label{eq:evoD}
    \\
    \left( \partial_t - \mathcal{L}_\beta \right) \mathcal{E} & = (\rho_E + p + \Pi) (\alpha v^iv^j K_{ij} - v^i \partial_i \alpha) \nonumber \\
    &\quad\quad - D_k \left[ \alpha v^k  \left( \mathcal{E} + p + \Pi\right) \right]   + \alpha K (\mathcal{E} + p + \Pi) ~,  \label{eq:evoE}
    \\
    \left( \partial_t - \mathcal{L}_\beta \right) S^i  &=    - D_k \left[ \alpha  \left( S^i v^k  + \gamma^{ik} (p+\Pi) \right) \right] - \rho_E D^i\alpha  + \alpha K S^i  ~, \label{eq:evoSi}
\end{align}
where $\mathcal{L}_\beta$ is the Lie derivative with respect to $\beta^i$, and $D_i$ is the covariant derivative compatible with the spatial metric $\gamma_{ij}$. Note that the bulk viscosity $\Pi$ contributes to the EoM for $S^i$ and is proportional to the fluid expansion $\Theta_\mathrm{f}$, which therefore needs to be evaluated explicitly. In principle, one can compute $\Theta_\mathrm{f}$ as long as the metric and the velocity field are given, and the calculation is greatly simplified by working in the fluid-comoving gauge $u^i=0$ by adopting a proper choice of $\alpha$ and $\beta^r$. One possible choice is to set $v^r=0$, $\beta^r=0$, and extract $\alpha$ from the time evolution of $v^r$, which is actually equivalent to the gauge choice used in the Misner-Sharp formalism~\cite{Misner:1964je}. In this case, the fluid four-velocity coincides with the unit normal vector to the constant-$t$ hyper-surface, i.e. $u^\mu = n^\mu$, and then $\Theta_\mathrm{f}$ can be evaluated as
\begin{align}\label{eq: Thetaf = -K}
    \Theta_\mathrm{f} = \nabla_\mu u^\mu = \nabla_\mu n^\mu = -K~.
\end{align}
Revisiting EoM for $S^i$~\eqref{eq:evoSi}, now it reduces to a purely spatial equation for the lapse function,
\begin{align} \label{eq: slicing condition}
    \partial_i \ln \alpha = -\frac{\partial_i(p+\Pi)}{\rho_E + p + \Pi}~,
\end{align}
which helps us to determine the slicing condition. The equation should be solved before each step of time evolution by setting the boundary condition $\alpha(r\to\infty)=1$ to match the background cosmology. The EoMs~\eqref{eq:evoD} and~\eqref{eq:evoE} originate from the particle current conservation and energy conservation, respectively~\cite{Alcubierre:2008book,Rezzolla:2013dea}, and summing up these two EoMs leads to the evolution equation for $\rho$ (equivalent to $\rho_E$ in the comoving gauge),
\begin{align}\label{eq: evoRho}
    \partial_t \rho = \alpha K (\rho + p + \Pi)~,
\end{align}
which serves as the evolution variable and is the only dynamical degree of freedom of the fluid.

Within this gauge, the EoMs for geometrical BSSN variables can be simplified. Comparing the EoMs for $a$ and $b$ with $\beta^r = 0$,
\begin{align}
    \partial_t \ln a = - 2\alpha A_a~, \quad \partial_t \ln b = - 2\alpha A_b~,
\end{align}
and using the traceless condition $A_a+2A_b =0$ and the boundary condition $a(r\to\infty)=b(r\to\infty)=1$, it can be shown that $a$ and $b$ are not independent, but are related through $a=1/b^2$. The independent evolution variables set now reduces to $\{\chi, a, K, A_a, \hat{\Delta}_r\}$. The first two variables $\chi$ and $a$ describe the spatial profile of the metric, followed by two variables $K$ and $A_a$ containing the time derivative information, as well as an auxiliary variable $\hat{\Delta}_r$ for numerical stability, the EoMs of which are given by
\begin{subequations}\label{eq: evoGeo}
\begin{align}
    \partial_t \chi &= -\frac{1}{6} \alpha K~,\\
    \partial_t a & = -2\alpha a A_a~,\\
    \partial_t K & = -\nabla^2\alpha + \alpha\left(\frac{3}{2}A_a^2 + \frac{1}{3}K^2 \right) + 4\pi \alpha (\rho + 3p + 3\Pi)~, \\
    \partial_t A_a & = -\left(\nabla^r\nabla_r\alpha - \frac{1}{3} \nabla^2\alpha \right) + \alpha \left({}^{(3)}\mathcal{R}_r^{\> r}-\frac{1}{3}{}^{(3)}\mathcal{R}\right) + \alpha K A_a~,\\
    \partial_t \hat{\Delta}^r & = -\frac{2}{a}\partial_r \left(\alpha A_a\right) + 2\alpha \left(A_a\hat{\Delta}_r  - \frac{2}{rb}(A_a - A_b)\right) + \frac{\eta\alpha}{a}\mathcal{M}_r~,
\end{align}
\end{subequations}
where $\mathcal{M}_r$ is the Momentum Constraint of the system (see definition below), and $\eta$ is an arbitrary parameter satisfying $\eta>1/2$~\cite{Alcubierre:2011pkc}, which is set to be $\eta=2$ in this work. The covariant derivatives of the lapse function are evaluated as
\begin{align}
    \nabla^2 \alpha &=  \frac{1}{a e^{4 \chi}} \left[ \partial_r^2 \alpha - \partial_r \alpha \left( \frac{\partial_r a}{a} - 2 \partial_r \chi - \frac{2}{r} \right) \right]~,\\
    \nabla^r \nabla_r \alpha &= \frac{1}{a e^{4 \chi}} \left[ \partial_r^2 \alpha - \partial_r \alpha \left( \frac{\partial_r a}{2a} + 2 \partial_r \chi \right) \right] ~.
\end{align}
${}^{(3)}\mathcal{R}_r^{\> r}$ and ${}^{(3)}\mathcal{R}$ are the $rr$-component of the spatial Ricci tensor and the spatial Ricci scalar, respectively, which are given by
\begin{align}
    {}^{(3)}\mathcal{R}^r_r = &- \frac{1}{a e^{4 \chi}} \biggl[ \frac{\partial^2_r a}{2a} 
    - a \partial_r \hat{\Delta}^r - \frac{3}{4} \left( \frac{\partial_r a}{a} \right)^2
    + \frac{1}{2} \left( \frac{\partial_r b}{b} \right)^2 - \frac{1}{2} \hat{\Delta}^r \partial_r a + \frac{\partial_r a}{rb}  \nonumber\\
    & + \frac{2}{r^2} \left( 1 - \frac{a}{b} \right) \left( 1 + \frac{r \partial_r b}{b} \right) + 4 \partial^2_r \chi - 2 \partial_r \chi \left( \frac{\partial_r a}{a} - \frac{\partial_r b}{b} - \frac{2}{r} \right) \biggr] ~, \\
    {}^{(3)}\mathcal{R} = &- \frac{1}{a e^{4 \chi}} \biggl[ \frac{\partial^2_r a}{2a} + \frac{\partial^2_r b}{b} - a \partial_r \hat{\Delta}^r - \left( \frac{\partial_r a}{a} \right)^2 + \frac{1}{2} \left( \frac{\partial_r b}{b} \right)^2 + \frac{2}{rb} \left( 3 - \frac{a}{b} \right) \partial_r b \nonumber\\
    &  + \frac{4}{r^2} \left( 1 - \frac{a}{b} \right) + 8 \left( \partial^2_r \chi + ( \partial_r \chi )^2 \right) - 8 \partial_r \chi \left( \frac{\partial_r a}{2a} - \frac{\partial_r b}{b} - \frac{2}{r} \right) \biggr] ~.  
\end{align}

There are four components in the Einstein equations that have no time derivatives and thus do not contribute to the dynamical evolutions. They are usually referred to as the Hamiltonian Constraint $\mathcal{H}$ and Momentum Constraints $\mathcal{M}_r$, and must be preserved under time evolution. In a spherical symmetric system, the non-vanishing constraints are given by
\begin{align} \label{eq: Ham}
    \mathcal{H} &= {}^{(3)}\mathcal{R} - K^2+K_{ij}K^{ij} - 16\pi \rho = {}^{(3)}\mathcal{R} - (A_a^2+2A_b^2) + \frac{2}{3}K^2 - 16\pi \rho~, \\
    \mathcal{M}_r & = \partial_r A_a - \frac{2}{3}\partial_r K + 6A_a \partial_r\chi + (A_a-A_b)\left(\frac{2}{r} + \frac{\partial_r b}{b}\right) - 8\pi S_r~.
    \label{eq: Mom}
\end{align}

\subsection{Apparent horizon}

The formation of a black hole is closely related to the formation of an apparent horizon, which is defined as the root of the expansions $\Theta_\pm$ to the null geodesics. Let us denote $l^\mu_\pm$ as the tangent vectors of the radially outgoing ($+$) and ingoing ($-$) null geodesics, respectively. Then the expansions $\Theta_\pm = \nabla_\mu l^\mu_\pm$ can be expressed in terms of BSSN variables as
\begin{align}
    \Theta_\pm = \pm \frac{\me^{-2\chi}}{\sqrt{a}} \left( 4\partial_r\chi + \frac{\partial_r b}{b} +\frac{2}{r} \right) +  A_a  - \frac{2}{3}K~.
\end{align}
Under ordinary circumstances, the null expansions on a $2$-dim sphere satisfy $\Theta_+>0$ and $\Theta_-<0$, reflecting the intuitive picture that ingoing null rays converge toward the interior, while outgoing null rays escape to the exterior. In the contexts of cosmology and black-hole physics, however, the expansions may change the sign because of the presence of the apparent horizon. In cosmological spacetimes, the apparent horizon is naturally associated with the Hubble horizon $R_\mathrm{H}$, outside of which the areal radius $R$ of an ingoing null ray keeps increasing, implying a positive expansion $\Theta_->0$. On the other hand, in black hole spacetimes, the apparent horizon is characterized by the vanishing of $\Theta_+$. To be more precise, a trapped region with $\Theta_+<0$ will form at some finite radius and then expand over time. The inner and outer boundaries of this region are denoted by $R_{\mathrm{BH},1}$ and $R_{\mathrm{BH},2}$, respectively. Therefore, for a PBH formation process in a cosmological background, the causal structure of null geodesics is described by
\begin{align}
\begin{aligned}
    \Theta_+>0 ~, ~\Theta_->0~, &\quad R>R_\mathrm{H}~, \\
    \Theta_+>0 ~, ~\Theta_-<0~, &\quad R_\mathrm{BH,2}<R<R_\mathrm{H}~, \\
    \Theta_+<0 ~, ~\Theta_-<0~, &\quad R_\mathrm{BH,1}<R<R_\mathrm{BH,2}~, \\
    \Theta_+<0 ~, ~\Theta_->0~, &\quad R<R_\mathrm{BH,1}~.
\end{aligned}
\end{align}

The mass of the central black hole is obtained from the Misner-Sharp-Hernandez (MSH) mass~\cite{Misner:1964je,Hernandez:1966zia} within an areal radius $R=\me^{2\chi}\sqrt{b} \> r$,
\begin{align}
    M = \frac{R}{2}\left(1 - g^{\mu\nu} \partial_\mu R \partial_\nu R\right)~.
\end{align}
This non-local mass is shown to be related to the product of two expansions by~\cite{Hayward:1993ph,Hayward:1994bu}
\begin{align}\label{eq: MSH mass from Theta}
    \frac{2M}{R} - 1 = \frac{R^2}{4}\Theta_+\Theta_-~.
\end{align}
Therefore, the black hole horizon position $R_\mathrm{BH,2}$ can be found numerically by searching for the larger root to $\Theta_+=0$, and the black hole mass can be extracted by evaluating the MSH mass at $R=R_\mathrm{BH,2}$.

\subsection{Initial data}

First of all, we specify the asymptotic spacetime at $r\to \infty$, i.e., the background cosmology, in terms of the BSSN variables. In the absence of perturbations, the background cosmology is described by an Friedmann–Lemaître–Robertson–Walker (FLRW) spacetime with EoS $p_\mathrm{bkg} = w_\mathrm{eff} \rho_\mathrm{bkg}$, where the evolutions of Hubble parameters and energy density are given by 
\begin{align}
    H_\mathrm{bkg} = \frac{2}{3(1+w_\mathrm{eff})} \frac{1}{t}~,\quad \rho_\mathrm{bkg} = \frac{3H^2_\mathrm{bkg}}{8\pi}~.
\end{align}

Next, we introduce curvature perturbation $\zeta(r)$ on top of the background metric, which is usually associated with a characteristic length scale $k_*^{-1}$. On the super-Hubble scale with $k_* \ll H$, the metric with curvature perturbation can be approximately written in a spatial isotropic coordinate as follows~\cite{Shibata:1999zs}
\begin{align}\label{eq: initial metric}
    \md s^2 = -\md t^2 + \mathfrak{a}(t)^2 \me^{2\zeta(r)} \left( \md r^2 + r ^2 \md\Omega^2 \right)~.
\end{align}
Consequently, the initial condition information that deviates from the background is all encoded in $\zeta(r)$. 
Given the statistical properties of the perturbations, the spatial profile $\zeta(r)$ can be characterized through the two-point correlation function of the associated random field in real space~\cite{Bardeen:1985tr}. In the simplest scenario, the perturbations are assumed to form a Gaussian random field, whose statistical properties are determined by the power spectrum of the scalar perturbation. For example, under the assumption of a narrow-peak power spectrum on small scales $\mathcal{P}_\zeta(k) \simeq A_\zeta \delta(k - k_*)$, the mean profile is given by a sinc function~\cite{Yoo:2018kvb,Yoo:2020dkz},
\begin{align}\label{eq: sinc}
    \zeta(r) = \mu^{(s)} \frac{\sin k_* r}{k_* r} \equiv \mu^{(s)} ~\mathrm{sinc}(k_*r)~.
\end{align}
Without specifying the detailed shape of the power spectrum, $\zeta(r)$ can also be modeled phenomenologically by a Gaussian function~\cite{Musco:2018rwt},
\begin{align}\label{eq: exp}
    \zeta(r) = \mu^{(g)} \exp\left(-(k_* r)^2\right)~.
\end{align}
More generally, a perturbation $\zeta_\mathrm{NG}$ with non-Gaussianity parameter $\gamma_\mathrm{NG}$ can be related to the underlying Gaussian random field by the logarithmic duality~\cite{Atal:2019cdz,Pi:2022ysn,Inui:2024sce,Inui:2024fgk},
\begin{align}
    \zeta_\mathrm{NG} = - \frac{1}{\gamma_\mathrm{NG}} \ln \left(1 - \gamma_\mathrm{NG} \zeta \right) ~. 
\end{align}
However, for simplicity, we only consider Gaussian curvature perturbations with the sinc initial profile~\eqref{eq: sinc} and the Gaussian initial profile~\eqref{eq: exp} in our numerical simulations.

The initial data are obtained by solving the constraint equations on the initial time slice.
We choose a conformally flat initial value with $a=b=1$, $A_a = A_b = \hat{\Delta}^r= 0$. Identifying the full metric~\eqref{eq: metric} with the ansatz~\eqref{eq: initial metric} gives $\chi = \zeta/2$. The extrinsic curvature is then fully described by its trace $K$, which is chosen to match the background $K=-3H_\mathrm{bkg}$. Solving the Momentum Constraints~\eqref{eq: Mom} directly gives $S_r=0$, which is consistent with the comoving gauge. Solving the Hamiltonian Constraint~\eqref{eq: Ham} results in
\begin{align}
    \rho &= \rho_\mathrm{bgk} + \delta\rho~, \\
    \delta\rho & = \frac{{}^{(3)}\mathcal{R}}{16\pi} = -\frac{\me^{-2\zeta}}{16\pi} \left( 4\zeta'' + \frac{8}{r}\zeta' + 2(\zeta')^2\right)~.
\end{align}

\subsection{Numerical implementation}

With the EoMs~\eqref{eq: evoRho} and~\eqref{eq: evoGeo} in hand, we now turn to the numerical implementation, which requires discretizing the system in both time and space. For the time direction, we adopt an explicit fourth-order Runge-Kutta (RK4) method to numerically integrate the system. For the spatial direction, we define a non-uniform radial grid $r$ in the interval $(r_\mathrm{min},r_\mathrm{max})$ with $N=4096$ cells, in terms of a uniform grid $x$ by $r = c_1 x + c_3 x^3$ with $0<x<1$, where the coefficients $c_1$ and $c_3$ can be uniquely determined by $r_\mathrm{min}$ and $r_\mathrm{max}$. The grid is cell-centred to avoid the $1/r$ divergence at $r=x=0$. The spatial derivative operators $\partial_r$ and $\partial_r^2$ are then related to $\partial_x$ and $\partial_x^2$ as follows,
\begin{align}
    \partial_r &= \frac{\md x}{\md r}\partial_x = (c_1+3c_3x^2)\partial_x~, \\
    \partial_r^2 &= \left(\frac{\md x}{\md r}\right)^2 \partial_x^2 + \frac{\md^2 x}{\md r^2}\partial_x = (c_1+3c_3x^2)^2\partial_x^2 + 6c_3x\partial_x~.
\end{align}
The derivatives on a uniform grid $x$ can be performed by the finite-difference method using a standard five-point stencil, providing fourth-order accuracy for the values, which is compatible with the RK4 method. We apply Kreiss-Oliger (KO) dissipations to suppress the high-frequency noise. For each evolution variable $f$ at the $i$-th cell, the KO term is added in the time derivatives~\cite{Palenzuela:2025ucx},
\begin{align}
    \partial_t f_i \to \partial_t f_i + \frac{\sigma}{64\Delta x} \left( f_{i-3} - 6f_{i-2} + 15f_{i-1} - 20f_{i} + 15f_{i+1} - 6f_{i+2} +f_{i+3} \right)~.
\end{align}

To complete the spatial discretization, boundary conditions must be specified, and ghost cells need to be introduced to evaluate spatial derivatives near the boundary. Because of the spherical symmetry of the system, reflective boundary conditions are applied to the inner boundary at $r\to 0$ by enforcing the appropriate parity. That is, fill in the ghost cell with $f_{-i}= f_i$ for variable $f$ with even parity, and $f_{-i} = -f_i$ for those with odd parity~\cite{Ruchlin:2017com}. In our case, only $\hat{\Delta}_r$ carries odd parity, while all the other variables have even parity. The outer boundary should match the background cosmology. Since $r_\mathrm{max}$ is chosen to be sufficiently large such that no perturbations propagate to the outer boundary within the simulation time, the spatial derivatives of all dynamical variables are expected to be negligible at the outer boundary. Consequently, we can simply impose a reflective boundary condition there to effectively mimic the background environment.

\section{Evolution of  Israel–Stewart viscosity}\label{app: IS}

Here we show how to extend the BSSN variables by one more variable $\Pi$ to describe a system by deriving the EoM for bulk viscosity in Israel–Stewart theory. We do not take any specific gauge choice in this section.

\subsection{Fluid expansion}

First of all, without assuming the evolution of viscosity, $\Pi$ should serve as an independent dynamical variable in the evolution. Therefore, the matter sector consists of four dynamical variables in total, $\{\mathcal{D, \mathcal{E}}, S_i, \Pi\}$. To the leading-order Eckart's theory, we know that $\Pi$ is closely related to the fluid expansion $\Theta_\mathrm{f}$. Projecting the covariant conservation of the stress tensor along the fluid's $4$-velocity, one can get
\begin{align}\label{eq: Theta_f from stress tensor}
    0 = u_\nu\nabla_\mu T^{\mu\nu} \quad \Rightarrow \quad \Theta_\mathrm{f} = -\frac{\partial_\lambda p}{(1+w)p+w\Pi}~,
\end{align}
where we have defined the derivative operator $\partial_\lambda \equiv u^\mu\partial_\mu$, and used the normalization $u_\mu u^\mu =-1$ and the EoS $p = w \rho$. Nevertheless, as the pressure $p$ is not evolved directly in our formalism, it should be expressed in terms of fundamental dynamical variables for further derivation. Let us introduce an auxiliary variable $z$ as
\begin{align}\label{eq: z def}
    z = (\rho+p+\Pi)W^2 = \mathcal{D}+\mathcal{E}+ p + \Pi~.
\end{align}
By applying the following equation
\begin{align}
    S^2 \equiv \gamma_{ij}S^i S^j = (\rho+p+\Pi)^2 W^4 v^2~,\quad v^2\equiv \gamma_{ij}v^iv^j~,
\end{align}
it can be found that
\begin{align}
    \rho+p+\Pi = \frac{z}{W^2} = z(1-v^2) = z \left(1 - \frac{S^2}{z^2}\right)~.
\end{align}
By using the EoS, the pressure can be reconstructed as
\begin{align}\label{eq: pressure of z}
    p = \frac{w}{1+w}(\rho + p) = \frac{w}{1+w}\left(z - \frac{S^2}{z} - \Pi \right)~.
\end{align}
Recall the definition for $z$ and replace $p$ with the equation above,
\begin{align}
    0 = z - (\mathcal{D}+\mathcal{E}+ p + \Pi) = \frac{z}{1+w} + \frac{wS^2}{(1+w)z} - \left(\mathcal{D}+\mathcal{E} + \frac{\Pi}{1+w}\right)~.
\end{align}
Solving the second-order equation for $z$ and substituting the solution into~\eqref{eq: pressure of z} leads to
\begin{align}\label{eq: p of BSSN}
    p = \frac{1}{2}\left( -(\mathcal{D}+\mathcal{E})(1-w) - \Pi + \sqrt{\left( (\mathcal{D}+\mathcal{E})(1+w)+ \Pi\right)^2 - 4 w S^2} ~\right)~.
\end{align}
In the inviscid limit $\Pi\to 0$, this result reduces back to the analytical relation found in previous literature~\cite{Staelens:2019sza}, where we have taken the plus branch of the solution that is valid for $0<w<1$. 

In principle, by substituting~\eqref{eq: p of BSSN} back to~\eqref{eq: Theta_f from stress tensor}, one can get a detailed expression for $\Theta_\mathrm{f}$, where the time derivative of $p$ should be translated to the BSSN variables by the chain rule. On the other hand, there is an alternative way to compute $\Theta_\mathrm{f}$, which comes from the conservation law for particle current rather than stress tensor~\cite{Alcubierre:2008book},
\begin{align}\label{eq: Theta_f from partical current}
    \nabla_\mu (\rho_0 u^\mu) = 0 \quad\Rightarrow\quad \Theta_\mathrm{f} = -\partial_\lambda \ln \rho_0 = - (\partial_\lambda \ln \mathcal{D} + \partial_\lambda \ln W)~.
\end{align}
Let us define the following abbreviations for later convenience,
\begin{align}
    f_1 = \mathcal{D} + \mathcal{E} + p + \Pi~,\quad f_2 = w(\rho + p + \Pi) = (1+w)p + w\Pi~.
\end{align}
Then, from Eq.~\eqref{eq: z def}, the Lorentz factor can be recast to
\begin{align}
    W^2 = \frac{\mathcal{D} + \mathcal{E} + p + \Pi}{\rho + p + \Pi} = w\frac{f_1}{f_2}~.
\end{align}
Plugging back into\eqref{eq: Theta_f from partical current}, the second term on the right-hand-sight can be re-written as
\begin{align}
    \partial_\lambda\ln W = \frac{1}{2}\left[ \frac{\partial_\lambda\mathcal{E}+ \partial_\lambda\mathcal{D}}{f_1} + \left(\frac{1}{f_1} - \frac{1+w}{f_2}\right)\partial_\lambda p + \left(\frac{1}{f_1} - \frac{w}{f_2}\right)\partial_\lambda \Pi \right]~.
\end{align}
In this way, we can cancel out the time derivative of $p$, which would otherwise require complicated applications of the chain rule, by performing a linear combination of~\eqref{eq: Theta_f from stress tensor} and~\eqref{eq: Theta_f from partical current}, yielding
\begin{align}
    &\frac{1}{2} \left( \frac{f_2}{f_1} - (1+w) \right)\cdot~\eqref{eq: Theta_f from stress tensor} +~\eqref{eq: Theta_f from partical current} \nonumber\\
    \Rightarrow\quad & \frac{X}{2f_1}\Theta_\mathrm{f} = -\partial_\lambda\ln\mathcal{D} + \frac{1}{2f_1} \left( \partial_\lambda\mathcal{E}+ \partial_\lambda\mathcal{D} - \frac{wS^2}{f_1 f_2}\partial_\lambda\Pi \right)
    \label{eq: Theta_f final}
\end{align}
with abbreviation
\begin{align}
    X \equiv 2p + \Pi + (1-w)(\mathcal{D}+\mathcal{E})=  \sqrt{\left( (\mathcal{D}+\mathcal{E})(1+w)+ \Pi\right)^2 - 4 w S^2}
\end{align}

\subsection{EoM of bulk viscosity}

The Israel–Stewart (IS) Theory~\cite{Israel:1976tn,Israel:1979wp} is a second-order, causal, and stable theory by introducing a relaxation time $\tau$,
\begin{align}\label{eq: IS viscosity}
    \tau \partial_\lambda \Pi + \Pi = -\xi\Theta_\mathrm{f} - \frac{1}{2}\tau \Pi\Theta_\mathrm{f} - S_\mathrm{higher~order}~.
\end{align}
As $\tau\to 0$, the equation above degenerates to Eckart's Theory $\Pi = -\xi\Theta_\mathrm{f}$. The higher-order contributions $S_\mathrm{higher~order}$ consist of time derivatives of logarithmic combinations of thermodynamic quantities and are consistently neglected here, the precise expression of which is given by~\cite{Ramirez:2020ldb,Nandi:2024jha}
\begin{align}
    S_\mathrm{higher~order} = \frac{1}{2}\tau \Pi \> \partial_\lambda \ln\frac{\tau}{\xi T}~.
\end{align}
Since $\Theta_\mathrm{f}$ also dependents on $\partial_\lambda \Pi$ according to~\eqref{eq: Theta_f final}, the EoM for $\Pi$ can be obtained from a rearrangement of~\eqref{eq: IS viscosity},
\begin{align}
    \left(\tau X - \left(\xi +\tau \frac{\Pi}{2}\right) \frac{wS^2}{f_1 f_2}\right) \partial_\lambda \Pi = -X \Pi - \left(\xi +\tau \frac{\Pi}{2}\right)\left(-2f_1\partial_\lambda \ln \mathcal{D} + \partial_\lambda\rho_E\right)
\end{align}
with $\rho_E = \mathcal{D} + \mathcal{E}$ for short. Writing the derivative operator explicitly in the coordinate we are working in,
\begin{align}
    \partial_\lambda = u^\mu \partial_\mu = \frac{W}{\alpha}\left(\partial_t - (\beta^i-\alpha v^i)\partial_i \right)~,
\end{align}
the last two terms on the right-hand-side of EoM can be further simplified through the EoMs~\eqref{eq:evoD} and~\eqref{eq:evoE},
\begin{align}
    \frac{\alpha}{W}\partial_\lambda\ln \mathcal{D} &= -D_i(\alpha v^i) + \alpha K~,  \\
    \frac{\alpha}{W}\ \partial_\lambda \rho_E &= f_1 (\alpha v^iv^j K_{ij} - v^i \partial_i \alpha) - \alpha v^i \partial_i ( p + \Pi ) - f_1 D_i(\alpha v^i) + \alpha K f_1 ~,  
\end{align}
which yields
\begin{align}
    -2f_1\partial_\lambda \ln \mathcal{D} + \partial_\lambda\rho_E = W\left( (D_iv^i - K + v^iv^j K_{ij} )f_1 - v^i\partial_i (p+\Pi)\right)~.
\end{align}
Therefore, the EoM reads
\begin{align}\label{eq: EoM for Pi}
    &\left(\tau X - \left(\xi +\tau \frac{\Pi}{2}\right) \frac{wS^2}{f_1 f_2}\right) \left(\partial_t \Pi -(\beta^i-\alpha v^i)\partial_i \Pi\right) \nonumber\\
     &\quad\quad = -\frac{\alpha X\Pi}{W} - \alpha \left(\xi +\tau \frac{\Pi}{2}\right)\left( (D_iv^i - K + v^iv^j K_{ij} )f_1 - v^i\partial_i (p+\Pi)\right)~.
\end{align}
In the comoving gauge with $v^i= 0$ and $\beta^i = 0$, the EoM reduces to
\begin{align}\label{eq: EoM for Pi comoving}
    \tau \partial_t\Pi = -\alpha(\Pi - \xi K) + \frac{1}{2} \alpha \tau \Pi K~.
\end{align}
If we further take Eckart's limit $\tau\to 0$, the EoM above gives out a constraint equation instead of an evolution equation, $\Pi - \xi K = 0$, which is equivalent to the combination of~\eqref{eq: Eckart viscosity} and~\eqref{eq: Thetaf = -K} we used in the main text.

It is noticed that, on the left-hand-side of~\eqref{eq: EoM for Pi} (as well as~\eqref{eq: EoM for Pi comoving}), the factors in front of $\partial_t\Pi$ such as $\tau$ and $S^2$ are very small, while on the right-hand-side the terms do not vanish in general, suggesting an explicit Runge-Kutta method no longer valid. To numerically solve the EoM, one should adopt other time-integration schemes to deal with the stiff terms properly, for example, the implicit-explicit Runge-Kutta (IMEX RK) method~\cite{Palenzuela:2025ucx}.

\begin{figure*}
    \centering
    \includegraphics[width=0.48\linewidth]{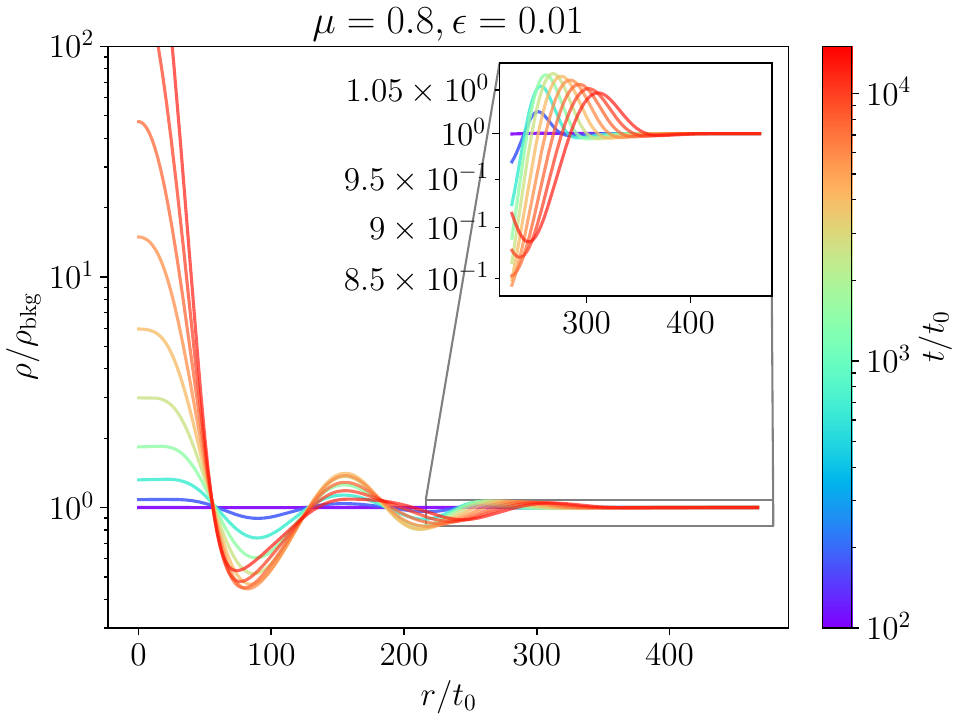} \quad
    \includegraphics[width=0.48\linewidth]{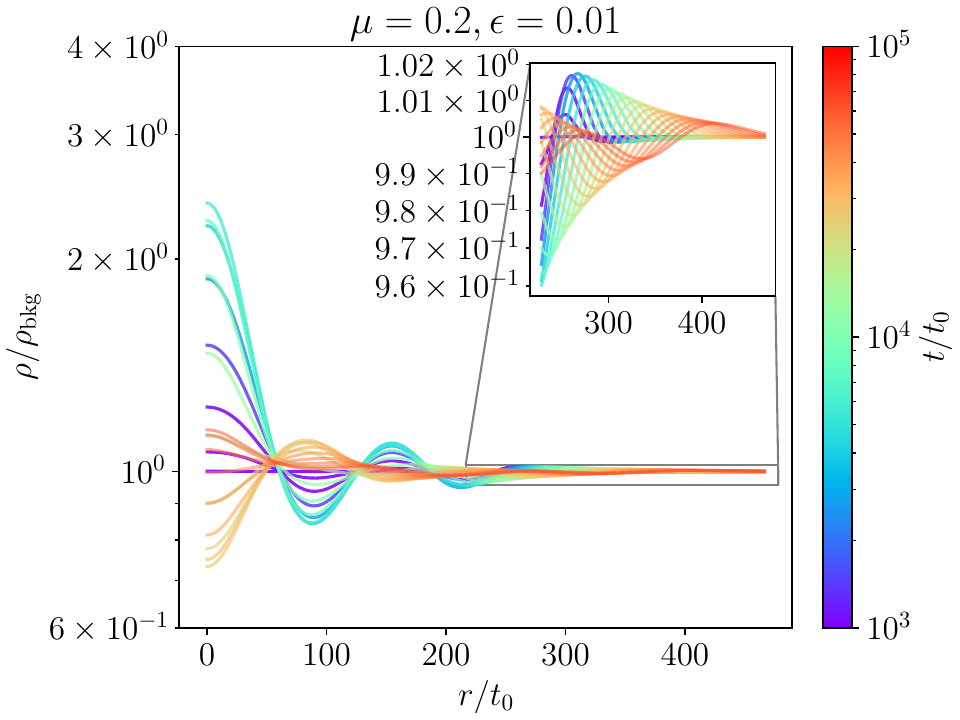} \\
    \vspace{0.3cm}
    \includegraphics[width=0.48\linewidth]{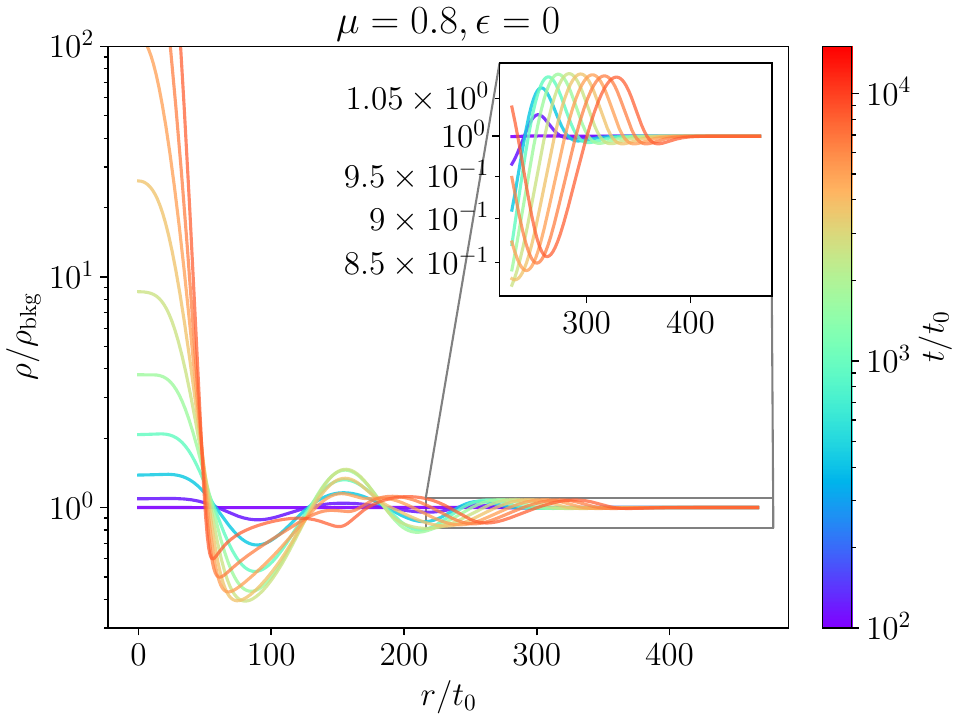} \quad
    \includegraphics[width=0.48\linewidth]{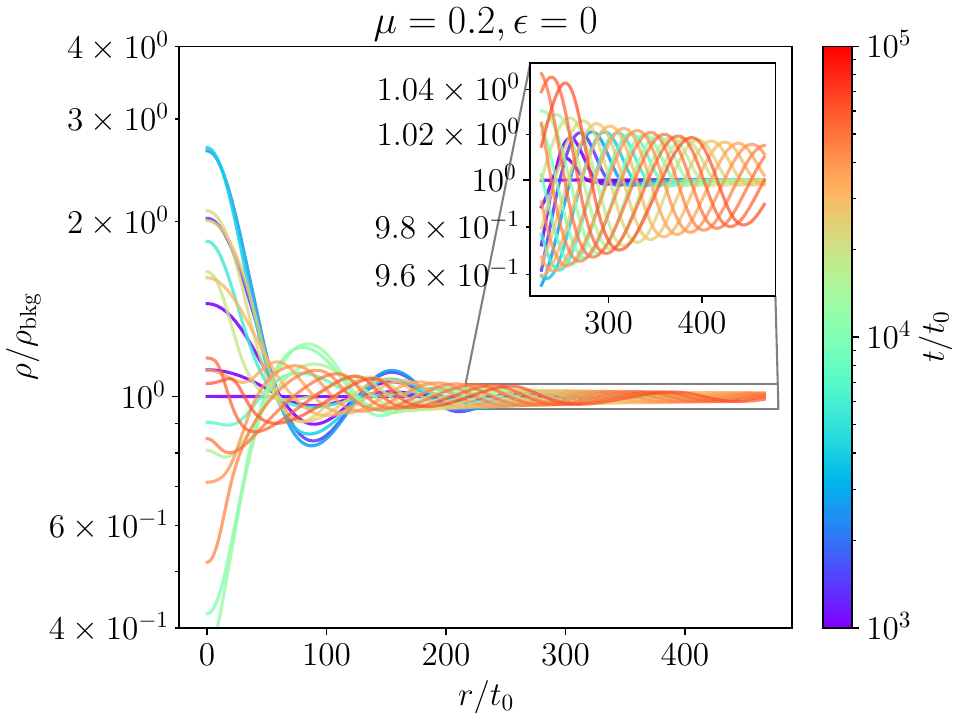}
    \caption{The evolution of spatial profiles for energy density ratio $\rho/\rho_\mathrm{bkg}$ with four parameter sets: $\mu=0.8$ (left), $0.2$ (right) and $\epsilon=0.01$ (top), $0$ (bottom), respectively. The EoS parameter is fixed as $w=1/3$.}
    \label{fig: dynamics}
\end{figure*}

\begin{figure*}
    \centering
    \includegraphics[width=0.47\linewidth]{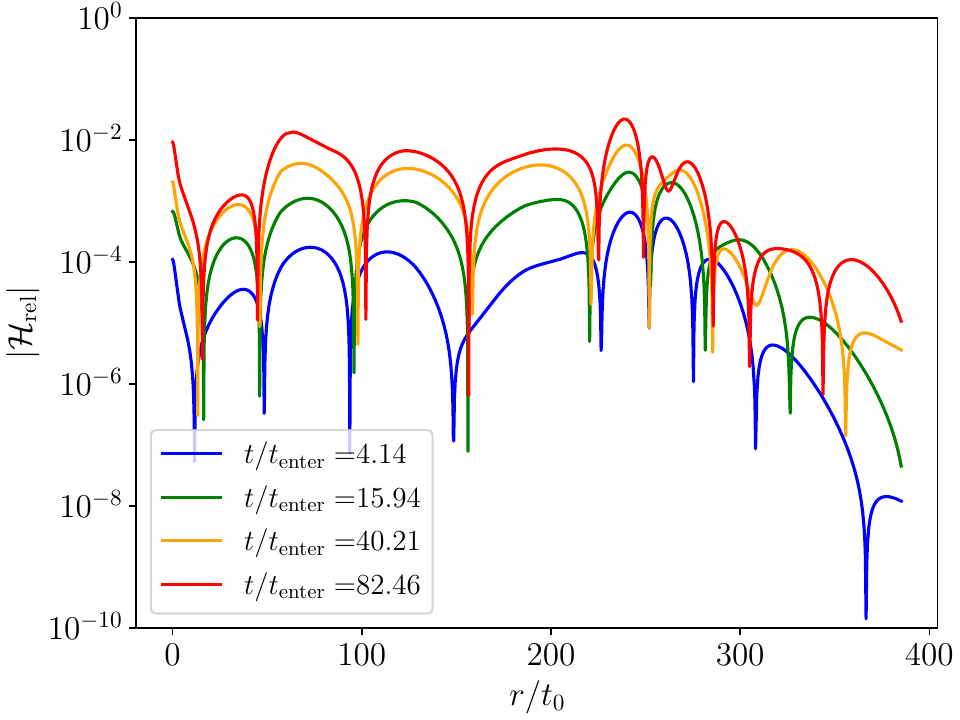} \quad
    \includegraphics[width=0.47\linewidth]{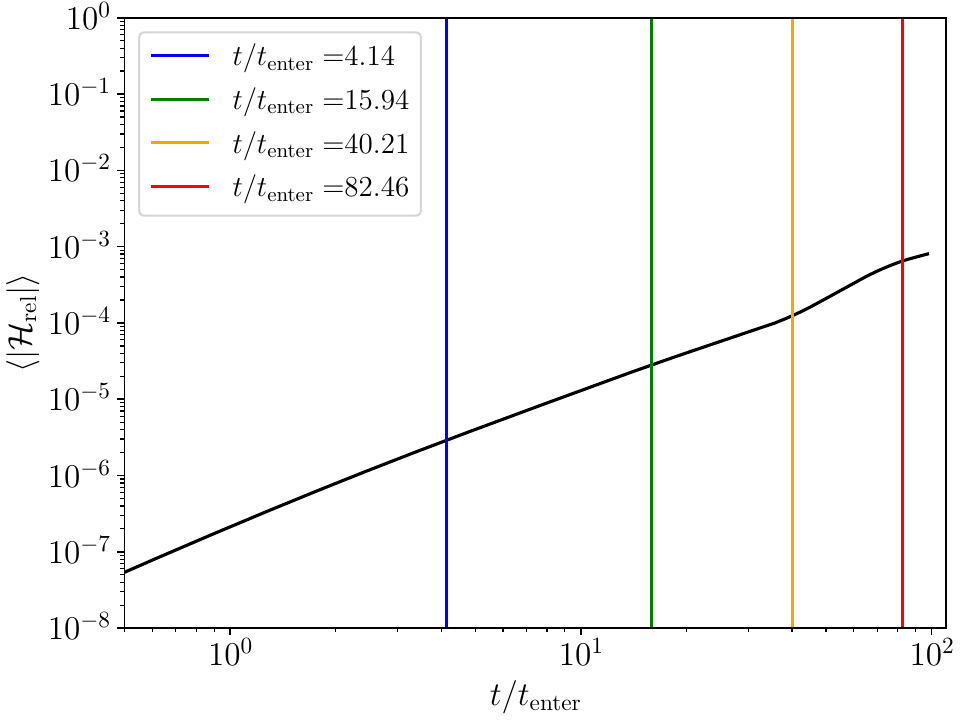}
    \caption{Left: The absolute value of relative violation of Hamiltonian constraint $|\mathcal{H}_\mathrm{rel}|$ at different time slices, where the parameter setting is $w=1/3$, $\mu=0.8$, $\epsilon=0.01$. Right: The grid-averaged violation $\langle|\mathcal{H}_\mathrm{rel}|\rangle$ at different time slices. The parameter is chosen to be the same as that in the left panel.}
    \label{fig: Hrel}
\end{figure*}

\section{Code validation}

In this section, we present our numerical results by solving the EoMs derived in the last section, for two kinds of initial profiles, Eqs.~\eqref{eq: sinc} and~\eqref{eq: exp}, respectively, with positive amplitudes $\mu$, corresponding to an overdense region at the centre. 

On the initial time slice, the perturbations are set to be well outside the Hubble volume. As they approach Hubble re-entry, the perturbations begin to grow nonlinearly due to the gravitational potential. The collapse into a black hole occurs if the initial amplitude of a perturbation $\mu$ exceeds a critical threshold $\mu_c$. In this case, the central overdense region will decouple from the cosmic expansion and form an apparent horizon within a finite time. Otherwise, the central overdensity will oscillate and finally disperse, converting its gravitational potential to the kinetic energy of the cosmic fluid and forming outgoing sound waves~\cite{Zeng:2025law,Ning:2025ogq,Ning:2025yvj}. An intuitive plot for the energy density normalized by the background $\rho/\rho_\mathrm{bkg}$ with typical parameter choices is presented in Fig.~\ref{fig: dynamics}.  In both collapsing  ($\mu=0.8$) and non-collapsing ($\mu=0.2$) cases, the presence of bulk viscosity noticeably delays the propagation of outgoing sound waves out of the potential well. Furthermore, viscous dissipative effects results in a faster attenuation of the sound wave amplitude compared to the inviscid case. 

At last, we show the validity of our simulations by checking the fulfillment of the constraint equations as customary. Specifically, we check the relative violation $\mathcal{H}_\mathrm{rel}$ of the Hamiltonian constraint defined as
\begin{align}
    \mathcal{H}_\mathrm{rel} \equiv 
    \frac{{}^{(3)}\mathcal{R} - K^2+K_{ij}K^{ij} - 16\pi \rho}
    { |{}^{(3)}\mathcal{R}| + |K^2| + |K_{ij}K^{ij}| + |16\pi \rho|}~.
\end{align}
A well-performed simulation should at least satisfy $|\mathcal{H}_\mathrm{rel}|\leq 10^{-2}$, while large violations deep inside the apparent horizon can be neglected as they do not affect the spacetime structure outside. Here we plot the spatial distribution and time evolution of $\mathcal{H}_\mathrm{rel}$ in Fig.~\ref{fig: Hrel}, where the time coordinate is normalized by the horizon-entering time $t_\mathrm{enter}$.

\bibliography{ref}

@misc{footnote,
      note = "See Supplemental Material for the simulation framework, evolution equations, and reuslts with validations,  which includes Refs.~\cite{Baumgarte:1998te,Shibata:1995we,Alcubierre:2011pkc,Eckart:1940,DiPrisco:2000dw,Carlevaro:2005xw,Maartens:1995wt,Israel:1976tn,Israel:1979wp,Alcubierre:2008book,Misner:1964je,Rezzolla:2013dea,Hernandez:1966zia,Hayward:1993ph,Hayward:1994bu,Shibata:1999zs,Bardeen:1985tr,Yoo:2018kvb,Yoo:2020dkz,Musco:2018rwt,Atal:2019cdz,Pi:2022ysn,Inui:2024sce,Inui:2024fgk,Palenzuela:2025ucx,Ruchlin:2017com,Staelens:2019sza,Ramirez:2020ldb,Nandi:2024jha,Palenzuela:2025ucx,Zeng:2025law,Ning:2025ogq,Ning:2025yvj}."
}

@article{Carr:2009jm,
    author = "Carr, B. J. and Kohri, Kazunori and Sendouda, Yuuiti and Yokoyama, Jun'ichi",
    title = "{New cosmological constraints on primordial black holes}",
    eprint = "0912.5297",
    archivePrefix = "arXiv",
    primaryClass = "astro-ph.CO",
    reportNumber = "RESCEU-31-09, TU-852, YITP-09-112",
    doi = "10.1103/PhysRevD.81.104019",
    journal = "Phys. Rev. D",
    volume = "81",
    pages = "104019",
    year = "2010"
}

@article{Carr:2020gox,
    author = "Carr, Bernard and Kohri, Kazunori and Sendouda, Yuuiti and Yokoyama, Jun'ichi",
    title = "{Constraints on primordial black holes}",
    eprint = "2002.12778",
    archivePrefix = "arXiv",
    primaryClass = "astro-ph.CO",
    reportNumber = "RESCEU-03/20; KEK-Cosmo-249; KEK-TH-2199; IPMU20-0024",
    doi = "10.1088/1361-6633/ac1e31",
    journal = "Rept. Prog. Phys.",
    volume = "84",
    number = "11",
    pages = "116902",
    year = "2021"
}

@article{Harada:2013epa,
    author = "Harada, Tomohiro and Yoo, Chul-Moon and Kohri, Kazunori",
    title = "{Threshold of primordial black hole formation}",
    eprint = "1309.4201",
    archivePrefix = "arXiv",
    primaryClass = "astro-ph.CO",
    reportNumber = "RUP-13-9, KEK-COSMO-129, KEK-TH-1668",
    doi = "10.1103/PhysRevD.88.084051",
    journal = "Phys. Rev. D",
    volume = "88",
    number = "8",
    pages = "084051",
    year = "2013",
    note = "[Erratum: Phys.Rev.D 89, 029903 (2014)]"
}

@article{Brevik:2017msy,
    author = "Brevik, Iver and Gr{\o}n, {\O}yvind and de Haro, Jaume and Odintsov, Sergei D. and Saridakis, Emmanuel N.",
    title = "{Viscous Cosmology for Early- and Late-Time Universe}",
    eprint = "1706.02543",
    archivePrefix = "arXiv",
    primaryClass = "gr-qc",
    doi = "10.1142/S0218271817300245",
    journal = "Int. J. Mod. Phys. D",
    volume = "26",
    number = "14",
    pages = "1730024",
    year = "2017"
}

@article{Niemeyer:1997mt,
    author = "Niemeyer, Jens C. and Jedamzik, K.",
    title = "{Near-critical gravitational collapse and the initial mass function of primordial black holes}",
    eprint = "astro-ph/9709072",
    archivePrefix = "arXiv",
    doi = "10.1103/PhysRevLett.80.5481",
    journal = "Phys. Rev. Lett.",
    volume = "80",
    pages = "5481--5484",
    year = "1998"
}

@article{LISACosmologyWorkingGroup:2023njw,
    author = "Bagui, Eleni and others",
    collaboration = "LISA Cosmology Working Group",
    title = "{Primordial black holes and their gravitational-wave signatures}",
    eprint = "2310.19857",
    archivePrefix = "arXiv",
    primaryClass = "astro-ph.CO",
    doi = "10.1007/s41114-024-00053-w",
    journal = "Living Rev. Rel.",
    volume = "28",
    number = "1",
    pages = "1",
    year = "2025"
}

@article{Escriva:2022duf,
    author = "Escriv{\`a}, Albert and Kuhnel, Florian and Tada, Yuichiro",
    editor = "Sedda, Manuel Arca and Bortolas, Elisa and Spera, Mario",
    title = "{Primordial Black Holes}",
    eprint = "2211.05767",
    archivePrefix = "arXiv",
    primaryClass = "astro-ph.CO",
    doi = "10.1016/B978-0-32-395636-9.00012-8",
    month = "11",
    year = "2022"
}

@article{Carr:2020xqk,
    author = "Carr, Bernard and Kuhnel, Florian",
    title = "{Primordial Black Holes as Dark Matter: Recent Developments}",
    eprint = "2006.02838",
    archivePrefix = "arXiv",
    primaryClass = "astro-ph.CO",
    doi = "10.1146/annurev-nucl-050520-125911",
    journal = "Ann. Rev. Nucl. Part. Sci.",
    volume = "70",
    pages = "355--394",
    year = "2020"
}

@article{Yoo:2022mzl,
    author = "Yoo, Chul-Moon",
    title = "{The Basics of Primordial Black Hole Formation and Abundance Estimation}",
    eprint = "2211.13512",
    archivePrefix = "arXiv",
    primaryClass = "astro-ph.CO",
    doi = "10.3390/galaxies10060112",
    journal = "Galaxies",
    volume = "10",
    number = "6",
    pages = "112",
    year = "2022"
}

@article{Joana:2025gqf,
    author = "Joana, Cristian and Yuwen, Zi-Yan",
    title = "{Primordial black holes from primordial voids}",
    eprint = "2510.11611",
    archivePrefix = "arXiv",
    primaryClass = "astro-ph.CO",
    doi = "10.1103/j3hw-d5cx",
    journal = "Phys. Rev. D",
    volume = "113",
    number = "2",
    pages = "023518",
    year = "2026"
}

@article{Zeldovich:1967lct,
    author = "Zel'dovich, Ya. B. and Novikov, I. D.",
    title = "{The Hypothesis of Cores Retarded during Expansion and the Hot Cosmological Model}",
    journal = "Sov. Astron.",
    volume = "10",
    pages = "602",
    year = "1967"
}

@article{Hawking:1971ei,
    author = "Hawking, Stephen",
    title = "{Gravitationally collapsed objects of very low mass}",
    doi = "10.1093/mnras/152.1.75",
    journal = "Mon. Not. Roy. Astron. Soc.",
    volume = "152",
    pages = "75",
    year = "1971"
}

@article{Carr:1974nx,
    author = "Carr, Bernard J. and Hawking, S. W.",
    title = "{Black holes in the early Universe}",
    doi = "10.1093/mnras/168.2.399",
    journal = "Mon. Not. Roy. Astron. Soc.",
    volume = "168",
    pages = "399--415",
    year = "1974"
}

@article{Carr:1975qj,
    author = "Carr, Bernard J.",
    title = "{The Primordial black hole mass spectrum}",
    doi = "10.1086/153853",
    journal = "Astrophys. J.",
    volume = "201",
    pages = "1--19",
    year = "1975"
}

@article{Carr:2016drx,
    author = "Carr, Bernard and Kuhnel, Florian and Sandstad, Marit",
    title = "{Primordial Black Holes as Dark Matter}",
    eprint = "1607.06077",
    archivePrefix = "arXiv",
    primaryClass = "astro-ph.CO",
    reportNumber = "NORDITA-2016-83",
    doi = "10.1103/PhysRevD.94.083504",
    journal = "Phys. Rev. D",
    volume = "94",
    number = "8",
    pages = "083504",
    year = "2016"
}

@article{Green:2020jor,
    author = "Green, Anne M. and Kavanagh, Bradley J.",
    title = "{Primordial Black Holes as a dark matter candidate}",
    eprint = "2007.10722",
    archivePrefix = "arXiv",
    primaryClass = "astro-ph.CO",
    doi = "10.1088/1361-6471/abc534",
    journal = "J. Phys. G",
    volume = "48",
    number = "4",
    pages = "043001",
    year = "2021"
}

@article{Carr:2021bzv,
    author = "Carr, Bernard and Kuhnel, Florian",
    title = "{Primordial black holes as dark matter candidates}",
    eprint = "2110.02821",
    archivePrefix = "arXiv",
    primaryClass = "astro-ph.CO",
    doi = "10.21468/SciPostPhysLectNotes.48",
    journal = "SciPost Phys. Lect. Notes",
    volume = "48",
    pages = "1",
    year = "2022"
}

@article{Niemeyer:1999ak,
    author = "Niemeyer, Jens C. and Jedamzik, K.",
    title = "{Dynamics of primordial black hole formation}",
    eprint = "astro-ph/9901292",
    archivePrefix = "arXiv",
    doi = "10.1103/PhysRevD.59.124013",
    journal = "Phys. Rev. D",
    volume = "59",
    pages = "124013",
    year = "1999"
}

@article{Passaglia:2021jla,
    author = "Passaglia, Samuel and Sasaki, Misao",
    title = "{Primordial black holes from CDM isocurvature perturbations}",
    eprint = "2109.12824",
    archivePrefix = "arXiv",
    primaryClass = "astro-ph.CO",
    reportNumber = "YITP-21-94",
    doi = "10.1103/PhysRevD.105.103530",
    journal = "Phys. Rev. D",
    volume = "105",
    number = "10",
    pages = "103530",
    year = "2022"
}

@article{Yoo:2021fxs,
    author = "Yoo, Chul-Moon and Harada, Tomohiro and Hirano, Shin'ichi and Okawa, Hirotada and Sasaki, Misao",
    title = "{Primordial black hole formation from massless scalar isocurvature}",
    eprint = "2112.12335",
    archivePrefix = "arXiv",
    primaryClass = "gr-qc",
    reportNumber = "YITP-21-161, RUP-21-23",
    doi = "10.1103/PhysRevD.105.103538",
    journal = "Phys. Rev. D",
    volume = "105",
    number = "10",
    pages = "103538",
    year = "2022"
}

@article{Green:2000he,
    author = "Green, Anne M. and Malik, Karim A.",
    title = "{Primordial black hole production due to preheating}",
    eprint = "hep-ph/0008113",
    archivePrefix = "arXiv",
    doi = "10.1103/PhysRevD.64.021301",
    journal = "Phys. Rev. D",
    volume = "64",
    pages = "021301",
    year = "2001"
}

@article{Jedamzik:2010dq,
    author = "Jedamzik, Karsten and Lemoine, Martin and Martin, Jerome",
    title = "{Collapse of Small-Scale Density Perturbations during Preheating in Single Field Inflation}",
    eprint = "1002.3039",
    archivePrefix = "arXiv",
    primaryClass = "astro-ph.CO",
    doi = "10.1088/1475-7516/2010/09/034",
    journal = "JCAP",
    volume = "09",
    pages = "034",
    year = "2010"
}

@article{Cotner:2018vug,
    author = "Cotner, Eric and Kusenko, Alexander and Takhistov, Volodymyr",
    title = "{Primordial Black Holes from Inflaton Fragmentation into Oscillons}",
    eprint = "1801.03321",
    archivePrefix = "arXiv",
    primaryClass = "astro-ph.CO",
    reportNumber = "IPMU18-0008",
    doi = "10.1103/PhysRevD.98.083513",
    journal = "Phys. Rev. D",
    volume = "98",
    number = "8",
    pages = "083513",
    year = "2018"
}

@article{Martin:2019nuw,
    author = "Martin, J{\'e}r{\^o}me and Papanikolaou, Theodoros and Vennin, Vincent",
    title = "{Primordial black holes from the preheating instability in single-field inflation}",
    eprint = "1907.04236",
    archivePrefix = "arXiv",
    primaryClass = "astro-ph.CO",
    doi = "10.1088/1475-7516/2020/01/024",
    journal = "JCAP",
    volume = "01",
    pages = "024",
    year = "2020"
}

@article{Kou:2019bbc,
    author = "Kou, Xiao-Xiao and Tian, Chi and Zhou, Shuang-Yong",
    title = "{Oscillon Preheating in Full General Relativity}",
    eprint = "1912.09658",
    archivePrefix = "arXiv",
    primaryClass = "gr-qc",
    doi = "10.1088/1361-6382/abd09f",
    journal = "Class. Quant. Grav.",
    volume = "38",
    number = "4",
    pages = "045005",
    year = "2021"
}

@article{Hawking:1982ga,
    author = "Hawking, S. W. and Moss, I. G. and Stewart, J. M.",
    title = "{Bubble Collisions in the Very Early Universe}",
    reportNumber = "Print-82-0180 (CAMBRIDGE)",
    doi = "10.1103/PhysRevD.26.2681",
    journal = "Phys. Rev. D",
    volume = "26",
    pages = "2681",
    year = "1982"
}

@article{Crawford:1982yz,
    author = "Crawford, Matt and Schramm, David N.",
    title = "{Spontaneous Generation of Density Perturbations in the Early Universe}",
    reportNumber = "EFI-82-04-CHICAGO",
    doi = "10.1038/298538a0",
    journal = "Nature",
    volume = "298",
    pages = "538--540",
    year = "1982"
}

@article{Moss:1994iq,
    author = "Moss, I. G.",
    title = "{Singularity formation from colliding bubbles}",
    doi = "10.1103/PhysRevD.50.676",
    journal = "Phys. Rev. D",
    volume = "50",
    pages = "676--681",
    year = "1994"
}

@article{Khlopov:1998nm,
    author = "Khlopov, M. Yu. and Konoplich, R. V. and Rubin, S. G. and Sakharov, A. S.",
    title = "{Formation of black holes in first order phase transitions}",
    eprint = "hep-ph/9807343",
    archivePrefix = "arXiv",
    reportNumber = "ROME1-1203-1998",
    month = "7",
    year = "1998"
}

@article{Khlopov:1999ys,
    author = "Khlopov, M. Yu. and Konoplich, R. V. and Rubin, S. G. and Sakharov, A. S.",
    editor = "Melnikov, V. N.",
    title = "{First order phase transitions as a source of black holes in the early universe}",
    eprint = "hep-ph/9912422",
    archivePrefix = "arXiv",
    journal = "Grav. Cosmol.",
    volume = "2",
    pages = "S1",
    year = "1999"
}

@article{Jung:2021mku,
    author = "Jung, Tae Hyun and Okui, Takemichi",
    title = "{Primordial black holes from bubble collisions during a first-order phase transition}",
    eprint = "2110.04271",
    archivePrefix = "arXiv",
    primaryClass = "hep-ph",
    reportNumber = "KEK-TH-2350",
    doi = "10.1103/PhysRevD.110.115014",
    journal = "Phys. Rev. D",
    volume = "110",
    number = "11",
    pages = "115014",
    year = "2024"
}

@article{Hawking:1987bn,
    author = "Hawking, S. W.",
    title = "{Black Holes From Cosmic Strings}",
    reportNumber = "Print-88-0310 (CAMBRIDGE)",
    doi = "10.1016/0370-2693(89)90206-2",
    journal = "Phys. Lett. B",
    volume = "231",
    pages = "237--239",
    year = "1989"
}

@article{Polnarev:1988dh,
    author = "Polnarev, Alexander and Zembowicz, Robert",
    title = "{Formation of Primordial Black Holes by Cosmic Strings}",
    reportNumber = "CAMK-194",
    doi = "10.1103/PhysRevD.43.1106",
    journal = "Phys. Rev. D",
    volume = "43",
    pages = "1106--1109",
    year = "1991"
}

@article{Garriga:1992nm,
    author = "Garriga, Jaume and Vilenkin, Alexander",
    title = "{Black holes from nucleating strings}",
    eprint = "hep-ph/9208212",
    archivePrefix = "arXiv",
    reportNumber = "TUTP-92-6, NSF-ITP-92-1161",
    doi = "10.1103/PhysRevD.47.3265",
    journal = "Phys. Rev. D",
    volume = "47",
    pages = "3265--3274",
    year = "1993"
}

@article{Widrow:1989vj,
    author = "Widrow, Lawrence M.",
    title = "{Dynamics of Thick Domain Walls}",
    reportNumber = "CFA-2878",
    doi = "10.1103/PhysRevD.40.1002",
    journal = "Phys. Rev. D",
    volume = "40",
    pages = "1002",
    year = "1989"
}

@article{Rubin:2000dq,
    author = "Rubin, S. G. and Khlopov, M. Yu. and Sakharov, A. S.",
    editor = "Khlopov, M. Yu. and Prokhorov, M. E. and Starobinsky, A. A.",
    title = "{Primordial black holes from nonequilibrium second order phase transition}",
    eprint = "hep-ph/0005271",
    archivePrefix = "arXiv",
    journal = "Grav. Cosmol.",
    volume = "6",
    pages = "51--58",
    year = "2000"
}

@article{Escriva:2024lmm,
    author = "Escriv{\`a}, Albert and Yoo, Chul-Moon",
    title = "{Simulations of ellipsoidal primordial black hole formation}",
    eprint = "2410.03452",
    archivePrefix = "arXiv",
    primaryClass = "gr-qc",
    doi = "10.1103/PhysRevD.112.083518",
    journal = "Phys. Rev. D",
    volume = "112",
    number = "8",
    pages = "083518",
    year = "2025"
}

@article{Escriva:2024aeo,
    author = "Escriv{\`a}, Albert and Yoo, Chul-Moon",
    title = "{Nonspherical effects on the mass function of primordial black holes}",
    eprint = "2410.03451",
    archivePrefix = "arXiv",
    primaryClass = "gr-qc",
    doi = "10.1103/4jbp-87wc",
    journal = "Phys. Rev. D",
    volume = "112",
    number = "8",
    pages = "L081304",
    year = "2025"
}

@article{Moss:1984zf,
    author = "Moss, I. G.",
    title = "{BLACK HOLE BUBBLES}",
    reportNumber = "Print-85-0005 (NEWCASTLE)",
    doi = "10.1103/PhysRevD.32.1333",
    journal = "Phys. Rev. D",
    volume = "32",
    pages = "1333",
    year = "1985"
}

@article{Hiscock:1987hn,
    author = "Hiscock, W. A.",
    title = "{CAN BLACK HOLES NUCLEATE VACUUM PHASE TRANSITIONS?}",
    doi = "10.1103/PhysRevD.35.1161",
    journal = "Phys. Rev. D",
    volume = "35",
    pages = "1161--1170",
    year = "1987"
}

@article{Jinno:2023vnr,
    author = "Jinno, Ryusuke and Kume, Jun'ya and Yamada, Masaki",
    title = "{Super-slow phase transition catalyzed by BHs and the birth of baby BHs}",
    eprint = "2310.06901",
    archivePrefix = "arXiv",
    primaryClass = "hep-ph",
    reportNumber = "TU-1209, RESCEU-18/23",
    doi = "10.1016/j.physletb.2024.138465",
    journal = "Phys. Lett. B",
    volume = "849",
    pages = "138465",
    year = "2024"
}

@article{Yuwen:2024gcf,
    author = "Yuwen, Zi-Yan and Joana, Cristian and Wang, Shao-Jiang and Cai, Rong-Gen",
    title = "{Bubbles kick off primordial black holes to form more binaries}",
    eprint = "2406.05838",
    archivePrefix = "arXiv",
    primaryClass = "gr-qc",
    doi = "10.1103/PhysRevResearch.7.023180",
    journal = "Phys. Rev. Res.",
    volume = "7",
    number = "2",
    pages = "023180",
    year = "2025"
}

@article{Wang:2025hwc,
    author = "Wang, Haonan and Zhang, Ying-li and Suyama, Teruaki",
    title = "{Nearly Monochromatic Primordial Black Holes as total Dark Matter from Bubble Collapse}",
    eprint = "2510.19233",
    archivePrefix = "arXiv",
    primaryClass = "astro-ph.CO",
    month = "10",
    year = "2025"
}

@article{Bugaev:2010bb,
    author = "Bugaev, Edgar and Klimai, Peter",
    title = "{Constraints on the induced gravitational wave background from primordial black holes}",
    eprint = "1012.4697",
    archivePrefix = "arXiv",
    primaryClass = "astro-ph.CO",
    doi = "10.1103/PhysRevD.83.083521",
    journal = "Phys. Rev. D",
    volume = "83",
    pages = "083521",
    year = "2011"
}

@article{Ferrante:2022mui,
    author = "Ferrante, Giacomo and Franciolini, Gabriele and Iovino, Junior., Antonio and Urbano, Alfredo",
    title = "{Primordial non-Gaussianity up to all orders: Theoretical aspects and implications for primordial black hole models}",
    eprint = "2211.01728",
    archivePrefix = "arXiv",
    primaryClass = "astro-ph.CO",
    doi = "10.1103/PhysRevD.107.043520",
    journal = "Phys. Rev. D",
    volume = "107",
    number = "4",
    pages = "043520",
    year = "2023"
}

@article{Zeng:2025law,
    author = "Zeng, Xiang-Xi and Ning, Zhuan and Yuwen, Zi-Yan and Wang, Shao-Jiang and Deng, Heling and Cai, Rong-Gen",
    title = "{Relic gravitational waves from primordial gravitational collapses}",
    eprint = "2504.11275",
    archivePrefix = "arXiv",
    primaryClass = "gr-qc",
    month = "4",
    year = "2025"
}

@article{Ning:2025ogq,
    author = "Ning, Zhuan and Zeng, Xiang-Xi and Yuwen, Zi-Yan and Wang, Shao-Jiang and Deng, Heling and Cai, Rong-Gen",
    title = "{Sound waves from primordial black hole formations}",
    eprint = "2504.12243",
    archivePrefix = "arXiv",
    primaryClass = "gr-qc",
    doi = "10.1103/2md2-pjv5",
    journal = "Phys. Rev. D",
    volume = "113",
    number = "2",
    pages = "024020",
    year = "2026"
}

@article{Ning:2025yvj,
    author = "Ning, Zhuan and Yuwen, Zi-Yan and Zeng, Xiang-Xi and Cai, Rong-Gen and Wang, Shao-Jiang",
    title = "{Acoustic gravitational waves from primordial curvature perturbations}",
    eprint = "2512.21151",
    archivePrefix = "arXiv",
    primaryClass = "gr-qc",
    month = "12",
    year = "2025"
}

@article{Ning:2026nfs,
    author = "Ning, Zhuan and Zeng, Xiang-Xi and Cai, Rong-Gen and Wang, Shao-Jiang",
    title = "{Numerical simulations of primordial black hole formation via delayed first-order phase transitions}",
    eprint = "2601.21878",
    archivePrefix = "arXiv",
    primaryClass = "gr-qc",
    month = "1",
    year = "2026"
}

@article{Clark:2018ghm,
    author = "Clark, Steven and Dutta, Bhaskar and Gao, Yu and Ma, Yin-Zhe and Strigari, Louis E.",
    title = "{21 cm limits on decaying dark matter and primordial black holes}",
    eprint = "1803.09390",
    archivePrefix = "arXiv",
    primaryClass = "astro-ph.HE",
    reportNumber = "MI-TH-1879",
    doi = "10.1103/PhysRevD.98.043006",
    journal = "Phys. Rev. D",
    volume = "98",
    number = "4",
    pages = "043006",
    year = "2018"
}

@article{Mena:2019nhm,
    author = "Mena, Olga and Palomares-Ruiz, Sergio and Villanueva-Domingo, Pablo and Witte, Samuel J.",
    title = "{Constraining the primordial black hole abundance with 21-cm cosmology}",
    eprint = "1906.07735",
    archivePrefix = "arXiv",
    primaryClass = "astro-ph.CO",
    doi = "10.1103/PhysRevD.100.043540",
    journal = "Phys. Rev. D",
    volume = "100",
    number = "4",
    pages = "043540",
    year = "2019"
}

@article{Zhao:2024jad,
    author = "Zhao, Meng-Lin and Wang, Sai and Zhang, Xin",
    title = "{Prospects for probing dark matter particles and primordial black holes with the Hongmeng mission using the 21 cm global spectrum at cosmic dawn}",
    eprint = "2412.19257",
    archivePrefix = "arXiv",
    primaryClass = "astro-ph.CO",
    doi = "10.1088/1475-7516/2025/07/039",
    journal = "JCAP",
    volume = "07",
    pages = "039",
    year = "2025"
}

@article{Zhao:2025ddy,
    author = "Zhao, Meng-Lin and Shao, Yue and Wang, Sai and Zhang, Xin",
    title = "{Prospects for probing dark matter particles and primordial black holes with the Square Kilometre Array using the 21 cm power spectrum at cosmic dawn*}",
    eprint = "2507.02651",
    archivePrefix = "arXiv",
    primaryClass = "astro-ph.CO",
    doi = "10.1088/1674-1137/ae1375",
    journal = "Chin. Phys.",
    volume = "50",
    number = "2",
    pages = "025101",
    year = "2026"
}

@article{Yin:2026hvw,
    author = "Yin, Ziwen and Cheng, Hanyu and Visinelli, Luca",
    title = "{Primordial Black Hole Abundance from Reionization}",
    eprint = "2602.06794",
    archivePrefix = "arXiv",
    primaryClass = "astro-ph.CO",
    month = "2",
    year = "2026"
}

@article{Griest:2013aaa,
    author = "Griest, Kim and Cieplak, Agnieszka M. and Lehner, Matthew J.",
    title = "{Experimental Limits on Primordial Black Hole Dark Matter from the First 2 yr of Kepler Data}",
    eprint = "1307.5798",
    archivePrefix = "arXiv",
    primaryClass = "astro-ph.CO",
    doi = "10.1088/0004-637X/786/2/158",
    journal = "Astrophys. J.",
    volume = "786",
    number = "2",
    pages = "158",
    year = "2014"
}

@article{Niikura:2017zjd,
    author = "Niikura, Hiroko and others",
    title = "{Microlensing constraints on primordial black holes with Subaru/HSC Andromeda observations}",
    eprint = "1701.02151",
    archivePrefix = "arXiv",
    primaryClass = "astro-ph.CO",
    doi = "10.1038/s41550-019-0723-1",
    journal = "Nature Astron.",
    volume = "3",
    number = "6",
    pages = "524--534",
    year = "2019"
}

@article{Niikura:2019kqi,
    author = "Niikura, Hiroko and Takada, Masahiro and Yokoyama, Shuichiro and Sumi, Takahiro and Masaki, Shogo",
    title = "{Constraints on Earth-mass primordial black holes from OGLE 5-year microlensing events}",
    eprint = "1901.07120",
    archivePrefix = "arXiv",
    primaryClass = "astro-ph.CO",
    doi = "10.1103/PhysRevD.99.083503",
    journal = "Phys. Rev. D",
    volume = "99",
    number = "8",
    pages = "083503",
    year = "2019"
}

@article{Blaineau:2022nhy,
    author = "Blaineau, T. and others",
    title = "{New limits from microlensing on Galactic black holes in the mass range 10 M{\ensuremath{\odot}} {\ensuremath{<}} M {\ensuremath{<}} 1000 M{\ensuremath{\odot}}}",
    eprint = "2202.13819",
    archivePrefix = "arXiv",
    primaryClass = "astro-ph.GA",
    doi = "10.1051/0004-6361/202243430",
    journal = "Astron. Astrophys.",
    volume = "664",
    pages = "A106",
    year = "2022"
}

@article{Oguri:2022fir,
    author = "Oguri, Masamune and Takhistov, Volodymyr and Kohri, Kazunori",
    title = "{Revealing dark matter dress of primordial black holes by cosmological lensing}",
    eprint = "2208.05957",
    archivePrefix = "arXiv",
    primaryClass = "astro-ph.CO",
    reportNumber = "IPMU22-0040, KEK-TH-2444, KEK-Cosmo-0293, KEK-QUP-2023-0031",
    doi = "10.1016/j.physletb.2023.138276",
    journal = "Phys. Lett. B",
    volume = "847",
    pages = "138276",
    year = "2023"
}

@article{Cai:2022kbp,
    author = "Cai, Rong-Gen and Chen, Tan and Wang, Shao-Jiang and Yang, Xing-Yu",
    title = "{Gravitational microlensing by dressed primordial black holes}",
    eprint = "2210.02078",
    archivePrefix = "arXiv",
    primaryClass = "astro-ph.CO",
    doi = "10.1088/1475-7516/2023/03/043",
    journal = "JCAP",
    volume = "03",
    pages = "043",
    year = "2023"
}

@article{Hernandez:1966zia,
    author = "Hernandez, Walter C. and Misner, Charles W.",
    title = "{Observer Time as a Coordinate in Relativistic Spherical Hydrodynamics}",
    doi = "10.1086/148525",
    journal = "Astrophys. J.",
    volume = "143",
    pages = "452",
    year = "1966"
}

@article{Hayward:1993ph,
    author = "Hayward, Sean A.",
    title = "{Quasilocal gravitational energy}",
    eprint = "gr-qc/9303030",
    archivePrefix = "arXiv",
    reportNumber = "PRINT-93-0302 (GARCHING)",
    doi = "10.1103/PhysRevD.49.831",
    journal = "Phys. Rev. D",
    volume = "49",
    pages = "831--839",
    year = "1994"
}

@article{Hayward:1994bu,
    author = "Hayward, Sean A.",
    title = "{Gravitational energy in spherical symmetry}",
    eprint = "gr-qc/9408002",
    archivePrefix = "arXiv",
    doi = "10.1103/PhysRevD.53.1938",
    journal = "Phys. Rev. D",
    pages = "1938--1949",
    year = "1996"
}

@article{Choptuik:1992jv,
    author = "Choptuik, Matthew W.",
    title = "{Universality and scaling in gravitational collapse of a massless scalar field}",
    reportNumber = "FPRINT-92-33",
    doi = "10.1103/PhysRevLett.70.9",
    journal = "Phys. Rev. Lett.",
    volume = "70",
    pages = "9--12",
    year = "1993"
}

@article{Evans:1994pj,
    author = "Evans, Charles R. and Coleman, Jason S.",
    title = "{Observation of critical phenomena and selfsimilarity in the gravitational collapse of radiation fluid}",
    eprint = "gr-qc/9402041",
    archivePrefix = "arXiv",
    reportNumber = "TAR-039-UNC",
    doi = "10.1103/PhysRevLett.72.1782",
    journal = "Phys. Rev. Lett.",
    volume = "72",
    pages = "1782--1785",
    year = "1994"
}

@article{Musco:2012au,
    author = "Musco, Ilia and Miller, John C.",
    title = "{Primordial black hole formation in the early universe: critical behaviour and self-similarity}",
    eprint = "1201.2379",
    archivePrefix = "arXiv",
    primaryClass = "gr-qc",
    doi = "10.1088/0264-9381/30/14/145009",
    journal = "Class. Quant. Grav.",
    volume = "30",
    pages = "145009",
    year = "2013"
}

@article{Shibata:1999zs,
    author = "Shibata, Masaru and Sasaki, Misao",
    title = "{Black hole formation in the Friedmann universe: Formulation and computation in numerical relativity}",
    eprint = "gr-qc/9905064",
    archivePrefix = "arXiv",
    reportNumber = "OU-TAP-93",
    doi = "10.1103/PhysRevD.60.084002",
    journal = "Phys. Rev. D",
    volume = "60",
    pages = "084002",
    year = "1999"
}

@article{Bardeen:1985tr,
    author = "Bardeen, James M. and Bond, J. R. and Kaiser, Nick and Szalay, A. S.",
    title = "{The Statistics of Peaks of Gaussian Random Fields}",
    reportNumber = "FERMILAB-PUB-85-148-A, NSF-ITP-85-93",
    doi = "10.1086/164143",
    journal = "Astrophys. J.",
    volume = "304",
    pages = "15--61",
    year = "1986"
}

@article{Yoo:2018kvb,
    author = "Yoo, Chul-Moon and Harada, Tomohiro and Garriga, Jaume and Kohri, Kazunori",
    title = "{Primordial black hole abundance from random Gaussian curvature perturbations and a local density threshold}",
    eprint = "1805.03946",
    archivePrefix = "arXiv",
    primaryClass = "astro-ph.CO",
    reportNumber = "RUP-18-15, KEK-Cosmo-225, KEK-TH-2052",
    doi = "10.1093/ptep/pty120",
    journal = "PTEP",
    volume = "2018",
    number = "12",
    pages = "123E01",
    year = "2018",
    note = "[Erratum: PTEP 2024, 049202 (2024)]"
}

@article{Yoo:2020dkz,
    author = "Yoo, Chul-Moon and Harada, Tomohiro and Hirano, Shin'ichi and Kohri, Kazunori",
    title = "{Abundance of Primordial Black Holes in Peak Theory for an Arbitrary Power Spectrum}",
    eprint = "2008.02425",
    archivePrefix = "arXiv",
    primaryClass = "astro-ph.CO",
    reportNumber = "RUP-20-25, KEK-Cosmo-261, KEK-TH-2245",
    doi = "10.1093/ptep/ptaa155",
    journal = "PTEP",
    volume = "2021",
    number = "1",
    pages = "013E02",
    year = "2021",
    note = "[Erratum: PTEP 2024, 049203 (2024)]"
}

@article{Musco:2018rwt,
    author = "Musco, Ilia",
    title = "{Threshold for primordial black holes: Dependence on the shape of the cosmological perturbations}",
    eprint = "1809.02127",
    archivePrefix = "arXiv",
    primaryClass = "gr-qc",
    doi = "10.1103/PhysRevD.100.123524",
    journal = "Phys. Rev. D",
    volume = "100",
    number = "12",
    pages = "123524",
    year = "2019"
}

@article{Atal:2019cdz,
    author = "Atal, Vicente and Garriga, Jaume and Marcos-Caballero, Airam",
    title = "{Primordial black hole formation with non-Gaussian curvature perturbations}",
    eprint = "1905.13202",
    archivePrefix = "arXiv",
    primaryClass = "astro-ph.CO",
    doi = "10.1088/1475-7516/2019/09/073",
    journal = "JCAP",
    volume = "09",
    pages = "073",
    year = "2019"
}

@article{Pi:2022ysn,
    author = "Pi, Shi and Sasaki, Misao",
    title = "{Logarithmic Duality of the Curvature Perturbation}",
    eprint = "2211.13932",
    archivePrefix = "arXiv",
    primaryClass = "astro-ph.CO",
    reportNumber = "IPMU22-0060, YITP-22-144",
    doi = "10.1103/PhysRevLett.131.011002",
    journal = "Phys. Rev. Lett.",
    volume = "131",
    number = "1",
    pages = "011002",
    year = "2023"
}

@article{Inui:2024sce,
    author = "Inui, Ryoto and Motohashi, Hayato and Pi, Shi and Tada, Yuichiro and Yokoyama, Shuichiro",
    title = "{Constant roll and non-Gaussian tail in light of logarithmic duality}",
    eprint = "2409.13500",
    archivePrefix = "arXiv",
    primaryClass = "astro-ph.CO",
    doi = "10.1088/1475-7516/2025/02/042",
    journal = "JCAP",
    volume = "02",
    pages = "042",
    year = "2025"
}

@article{Inui:2024fgk,
    author = "Inui, Ryoto and Joana, Cristian and Motohashi, Hayato and Pi, Shi and Tada, Yuichiro and Yokoyama, Shuichiro",
    title = "{Primordial black holes and induced gravitational waves from logarithmic non-Gaussianity}",
    eprint = "2411.07647",
    archivePrefix = "arXiv",
    primaryClass = "astro-ph.CO",
    doi = "10.1088/1475-7516/2025/03/021",
    journal = "JCAP",
    volume = "03",
    pages = "021",
    year = "2025"
}

@article{Misner:1964je,
    author = "Misner, Charles W. and Sharp, David H.",
    title = "{Relativistic equations for adiabatic, spherically symmetric gravitational collapse}",
    doi = "10.1103/PhysRev.136.B571",
    journal = "Phys. Rev.",
    volume = "136",
    pages = "B571--B576",
    year = "1964"
}

@article{Baumgarte:1998te,
    author = "Baumgarte, Thomas W. and Shapiro, Stuart L.",
    title = "{On the numerical integration of Einstein's field equations}",
    eprint = "gr-qc/9810065",
    archivePrefix = "arXiv",
    doi = "10.1103/PhysRevD.59.024007",
    journal = "Phys. Rev. D",
    volume = "59",
    pages = "024007",
    year = "1998"
}

@article{Shibata:1995we,
    author = "Shibata, Masaru and Nakamura, Takashi",
    title = "{Evolution of three-dimensional gravitational waves: Harmonic slicing case}",
    doi = "10.1103/PhysRevD.52.5428",
    journal = "Phys. Rev. D",
    volume = "52",
    pages = "5428--5444",
    year = "1995"
}

@article{Alcubierre:2011pkc,
    author = "Alcubierre, Miguel and Mendez, Martha D.",
    title = "{Formulations of the 3+1 evolution equations in curvilinear coordinates}",
    eprint = "1010.4013",
    archivePrefix = "arXiv",
    primaryClass = "gr-qc",
    doi = "10.1007/s10714-011-1202-x",
    journal = "Gen. Rel. Grav.",
    volume = "43",
    pages = "2769--2806",
    year = "2011"
}

@article{Palenzuela:2025ucx,
    author = "Palenzuela, Carlos and others",
    title = "{MHDuet: a high-order general relativistic radiation MHD code for CPU and GPU architectures}",
    eprint = "2510.13965",
    archivePrefix = "arXiv",
    primaryClass = "gr-qc",
    doi = "10.1088/1361-6382/ae255e",
    journal = "Class. Quant. Grav.",
    volume = "42",
    number = "24",
    pages = "245005",
    year = "2025"
}

@article{Ruchlin:2017com,
    author = "Ruchlin, Ian and Etienne, Zachariah B. and Baumgarte, Thomas W.",
    title = "{SENR/NRPy+: Numerical Relativity in Singular Curvilinear Coordinate Systems}",
    eprint = "1712.07658",
    archivePrefix = "arXiv",
    primaryClass = "gr-qc",
    doi = "10.1103/PhysRevD.97.064036",
    journal = "Phys. Rev. D",
    volume = "97",
    number = "6",
    pages = "064036",
    year = "2018"
}

@article{Baumann:2010tm,
    author = "Baumann, Daniel and Nicolis, Alberto and Senatore, Leonardo and Zaldarriaga, Matias",
    title = "{Cosmological Non-Linearities as an Effective Fluid}",
    eprint = "1004.2488",
    archivePrefix = "arXiv",
    primaryClass = "astro-ph.CO",
    doi = "10.1088/1475-7516/2012/07/051",
    journal = "JCAP",
    volume = "07",
    pages = "051",
    year = "2012"
}

@article{Ballesteros:2011cm,
    author = "Ballesteros, Guillermo and Hollenstein, Lukas and Jain, Rajeev Kumar and Kunz, Martin",
    title = "{Nonlinear cosmological consistency relations and effective matter stresses}",
    eprint = "1112.4837",
    archivePrefix = "arXiv",
    primaryClass = "astro-ph.CO",
    doi = "10.1088/1475-7516/2012/05/038",
    journal = "JCAP",
    volume = "05",
    pages = "038",
    year = "2012"
}

@article{Giovannini:2015uia,
    author = "Giovannini, Massimo",
    title = "{Non-linear curvature inhomogeneities and backreaction for relativistic viscous fluids}",
    eprint = "1503.08739",
    archivePrefix = "arXiv",
    primaryClass = "hep-th",
    reportNumber = "CERN-PH-TH-2015-068",
    doi = "10.1088/0264-9381/32/15/155004",
    journal = "Class. Quant. Grav.",
    volume = "32",
    pages = "155004",
    year = "2015"
}

@article{Maartens:1995wt,
    author = "Maartens, Roy",
    title = "{Dissipative cosmology}",
    doi = "10.1088/0264-9381/12/6/011",
    journal = "Class. Quant. Grav.",
    volume = "12",
    pages = "1455--1465",
    year = "1995"
}

@article{Maartens:1996dk,
    author = "Maartens, Roy and Mendez, Vicenc",
    title = "{Nonlinear bulk viscosity and inflation}",
    eprint = "astro-ph/9611205",
    archivePrefix = "arXiv",
    doi = "10.1103/PhysRevD.55.1937",
    journal = "Phys. Rev. D",
    volume = "55",
    pages = "1937--1942",
    year = "1997"
}

@article{Zimdahl:1996ka,
    author = "Zimdahl, Winfried",
    title = "{Bulk viscous cosmology}",
    eprint = "astro-ph/9601189",
    archivePrefix = "arXiv",
    doi = "10.1103/PhysRevD.53.5483",
    journal = "Phys. Rev. D",
    volume = "53",
    pages = "5483--5493",
    year = "1996"
}

@article{Bemfica:2020zjp,
    author = "Bemfica, Fabio S. and Disconzi, Marcelo M. and Noronha, Jorge",
    title = "{First-Order General-Relativistic Viscous Fluid Dynamics}",
    eprint = "2009.11388",
    archivePrefix = "arXiv",
    primaryClass = "gr-qc",
    doi = "10.1103/PhysRevX.12.021044",
    journal = "Phys. Rev. X",
    volume = "12",
    number = "2",
    pages = "021044",
    year = "2022"
}

@article{Singh:2011dw,
    author = "Singh, C. P.",
    title = "{Viscous FRW model with particle creation in the early universe}",
    eprint = "1107.2212",
    archivePrefix = "arXiv",
    primaryClass = "gr-qc",
    doi = "10.1142/S0217732312500708",
    journal = "Mod. Phys. Lett. A",
    volume = "27",
    pages = "1250070",
    year = "2012"
}

@article{Eshaghi:2015tqa,
    author = "Eshaghi, Mehdi and Riazi, Nematollah and Kiasatpour, Ahmad",
    title = "{Bulk Viscosity and Particle Creation in the Inflationary Cosmology}",
    eprint = "1504.07774",
    archivePrefix = "arXiv",
    primaryClass = "gr-qc",
    month = "4",
    year = "2015"
}

@article{Paul:2025rqe,
    author = "Paul, Tanmoy",
    title = "{Origin of bulk viscosity in cosmology and its thermodynamic implications}",
    eprint = "2504.00422",
    archivePrefix = "arXiv",
    primaryClass = "gr-qc",
    doi = "10.1103/PhysRevD.111.083540",
    journal = "Phys. Rev. D",
    volume = "111",
    number = "8",
    pages = "083540",
    year = "2025"
}

@article{Buoninfante:2016ixe,
    author = "Buoninfante, L. and Lambiase, G.",
    title = "{Cosmology with bulk viscosity and the gravitino problem}",
    eprint = "1610.01827",
    archivePrefix = "arXiv",
    primaryClass = "astro-ph.CO",
    doi = "10.1140/epjc/s10052-017-4840-7",
    journal = "Eur. Phys. J. C",
    volume = "77",
    number = "5",
    pages = "287",
    year = "2017"
}

@article{Eckart:1940,
  title = {The Thermodynamics of Irreversible Processes. III. Relativistic Theory of the Simple Fluid},
  author = {Eckart, Carl},
  journal = {Phys. Rev.},
  volume = {58},
  issue = {10},
  pages = {919--924},
  numpages = {0},
  year = {1940},
  month = {Nov},
  publisher = {American Physical Society},
  doi = {10.1103/PhysRev.58.919},
  url = {https://link.aps.org/doi/10.1103/PhysRev.58.919}
}

@article{DiPrisco:2000dw,
    author = "Di Prisco, A. and Herrera, L. and Ibanez, J.",
    title = "{Qualitative analysis of dissipative cosmologies}",
    eprint = "gr-qc/0010021",
    archivePrefix = "arXiv",
    doi = "10.1103/PhysRevD.63.023501",
    journal = "Phys. Rev. D",
    volume = "63",
    pages = "023501",
    year = "2001"
}

@article{Carlevaro:2005xw,
    author = "Carlevaro, Nakia and Montani, Giovanni",
    title = "{Bulk viscosity effects on the early universe stability}",
    eprint = "gr-qc/0506044",
    archivePrefix = "arXiv",
    doi = "10.1142/S0217732305017998",
    journal = "Mod. Phys. Lett. A",
    volume = "20",
    pages = "1729--1740",
    year = "2005"
}

@article{Israel:1976tn,
    author = "Israel, W.",
    title = "{Nonstationary irreversible thermodynamics: A Causal relativistic theory}",
    doi = "10.1016/0003-4916(76)90064-6",
    journal = "Annals Phys.",
    volume = "100",
    pages = "310--331",
    year = "1976"
}

@article{Israel:1979wp,
    author = "Israel, W. and Stewart, J. M.",
    title = "{Transient relativistic thermodynamics and kinetic theory}",
    doi = "10.1016/0003-4916(79)90130-1",
    journal = "Annals Phys.",
    volume = "118",
    pages = "341--372",
    year = "1979"
}

@article{Ramirez:2020ldb,
    author = "Ram{\'\i}rez, David Alejandro Tamayo",
    title = "{Thermodynamics of viscous dark energy for the late future time universe}",
    eprint = "2006.14153",
    archivePrefix = "arXiv",
    primaryClass = "gr-qc",
    doi = "10.31349/RevMexFis.68.020704",
    journal = "Rev. Mex. Fis.",
    volume = "68",
    number = "2",
    pages = "020704",
    year = "2022"
}

@article{Nandi:2024jha,
    author = "Nandi, Tanmay and Choudhuri, Amitava",
    title = "{Symmetry-based study and dynamics of casual bulk viscous matter-dominated universe}",
    doi = "10.1140/epjc/s10052-024-12637-5",
    journal = "Eur. Phys. J. C",
    volume = "84",
    number = "3",
    pages = "336",
    year = "2024"
}

@book{Alcubierre:2008book,
    author = {Alcubierre, Miguel},
    title = {Introduction to 3+1 Numerical Relativity},
    publisher = {Oxford University Press},
    year = {2008},
    month = {04},
    isbn = {9780199205677},
    doi = {10.1093/acprof:oso/9780199205677.001.0001},
    url = {https://doi.org/10.1093/acprof:oso/9780199205677.001.0001},
}

@book{Rezzolla:2013dea,
    author = "Rezzolla, Luciano and Zanotti, Olindo",
    title = "{Relativistic Hydrodynamics}",
    doi = "10.1093/acprof:oso/9780198528906.001.0001",
    isbn = "978-0-19-174650-5, 978-0-19-852890-6",
    publisher = "Oxford University Press",
    month = "9",
    year = "2013"
}

@article{Staelens:2019sza,
    author = {Staelens, Fran{\c{c}}ois and Rekier, J{\'e}r{\'e}my and F{\"u}zfa, Andr{\'e}},
    title = "{Universality of the spherical collapse with respect to the matter type : the case of a barotropic fluid with linear equation of state}",
    eprint = "1912.00677",
    archivePrefix = "arXiv",
    primaryClass = "gr-qc",
    doi = "10.1007/s10714-021-02804-4",
    journal = "Gen. Rel. Grav.",
    volume = "53",
    number = "4",
    pages = "38",
    year = "2021"
}

@article{Germani:2025hcu,
    author = "Germani, Cristiano and Montell{\`a}, Laia",
    title = "{The trichotomy of primordial black holes initial conditions}",
    eprint = "2510.02006",
    archivePrefix = "arXiv",
    primaryClass = "gr-qc",
    month = "10",
    year = "2025"
}

@article{Avelino:2013mua,
    author = "Avelino, Arturo and Garcia-Salcedo, Ricardo and Gonzalez, Tame and Nucamendi, Ulises and Quiros, Israel",
    title = "{Bulk Viscous Matter-dominated Universes: Asymptotic Properties}",
    eprint = "1303.5167",
    archivePrefix = "arXiv",
    primaryClass = "gr-qc",
    doi = "10.1088/1475-7516/2013/08/012",
    journal = "JCAP",
    volume = "08",
    pages = "012",
    year = "2013"
}

@article{Avelino:2013wea,
    author = "Avelino, Arturo and Leyva, Yoelsy and Urena-Lopez, L. Arturo",
    title = "{Interacting viscous dark fluids}",
    eprint = "1306.3270",
    archivePrefix = "arXiv",
    primaryClass = "astro-ph.CO",
    doi = "10.1103/PhysRevD.88.123004",
    journal = "Phys. Rev. D",
    volume = "88",
    pages = "123004",
    year = "2013"
}

@article{Normann:2016zby,
    author = "Normann, Ben David and Brevik, Iver",
    title = "{Characteristic Properties of Two Different Viscous Cosmology Models for the Future Universe}",
    eprint = "1612.01794",
    archivePrefix = "arXiv",
    primaryClass = "gr-qc",
    doi = "10.1142/S0217732317500262",
    journal = "Mod. Phys. Lett. A",
    volume = "32",
    number = "4",
    pages = "1750026",
    year = "2017"
}

@article{Normann:2021bjy,
    author = "Normann, Ben David and Brevik, Iver H{\r{a}}kon",
    title = "{Can the Hubble tension be resolved by bulk viscosity?}",
    eprint = "2107.13533",
    archivePrefix = "arXiv",
    primaryClass = "gr-qc",
    doi = "10.1142/S0217732321501984",
    journal = "Mod. Phys. Lett. A",
    volume = "36",
    number = "27",
    pages = "2150198",
    year = "2021"
}

@article{Dong:2015yjs,
    author = "Dong, Ruifeng and Kinney, William H and Stojkovic, Dejan",
    title = "{Gravitational wave production by Hawking radiation from rotating primordial black holes}",
    eprint = "1511.05642",
    archivePrefix = "arXiv",
    primaryClass = "astro-ph.CO",
    doi = "10.1088/1475-7516/2016/10/034",
    journal = "JCAP",
    volume = "10",
    pages = "034",
    year = "2016"
}

@article{Garcia-Bellido:1996mdl,
archiveprefix = {arXiv},
author = {Garcia-Bellido, Juan and Linde, Andrei D. and Wands, David},
doi = {10.1103/PhysRevD.54.6040},
eprint = {astro-ph/9605094},
journal = {Phys. Rev. D},
pages = {6040--6058},
reportnumber = {SU-ITP-96-20, SUSSEX-AST-96-5-1},
title = {{Density perturbations and black hole formation in hybrid inflation}},
volume = {54},
year = {1996}}

@article{Alabidi:2009bk,
archiveprefix = {arXiv},
author = {Alabidi, Laila and Kohri, Kazunori},
doi = {10.1103/PhysRevD.80.063511},
eprint = {0906.1398},
journal = {Phys. Rev. D},
pages = {063511},
primaryclass = {astro-ph.CO},
title = {{Generating Primordial Black Holes Via Hilltop-Type Inflation Models}},
volume = {80},
year = {2009}}

@article{Kawasaki:2012wr,
archiveprefix = {arXiv},
author = {Kawasaki, Masahiro and Kitajima, Naoya and Yanagida, Tsutomu T.},
doi = {10.1103/PhysRevD.87.063519},
eprint = {1207.2550},
journal = {Phys. Rev. D},
number = {6},
pages = {063519},
primaryclass = {hep-ph},
reportnumber = {ICRR-REPORT-616-2012-5, IPMU-12-0116},
title = {{Primordial black hole formation from an axionlike curvaton model}},
volume = {87},
year = {2013}}

@article{Clesse:2015wea,
archiveprefix = {arXiv},
author = {Clesse, S\'ebastien and Garc\'\i{}a-Bellido, Juan},
doi = {10.1103/PhysRevD.92.023524},
eprint = {1501.07565},
journal = {Phys. Rev. D},
number = {2},
pages = {023524},
primaryclass = {astro-ph.CO},
title = {{Massive Primordial Black Holes from Hybrid Inflation as Dark Matter and the seeds of Galaxies}},
volume = {92},
year = {2015}}

@article{Garcia-Bellido:2016dkw,
archiveprefix = {arXiv},
author = {Garcia-Bellido, Juan and Peloso, Marco and Unal, Caner},
doi = {10.1088/1475-7516/2016/12/031},
eprint = {1610.03763},
journal = {JCAP},
pages = {031},
primaryclass = {astro-ph.CO},
reportnumber = {IFT-UAM-CSIC-16-100, UMN-TH-3607-16},
title = {{Gravitational waves at interferometer scales and primordial black holes in axion inflation}},
volume = {12},
year = {2016}}

@article{Carr:2017edp,
archiveprefix = {arXiv},
author = {Carr, Bernard and Tenkanen, Tommi and Vaskonen, Ville},
doi = {10.1103/PhysRevD.96.063507},
eprint = {1706.03746},
journal = {Phys. Rev. D},
number = {6},
pages = {063507},
primaryclass = {astro-ph.CO},
title = {{Primordial black holes from inflaton and spectator field perturbations in a matter-dominated era}},
volume = {96},
year = {2017}}

@article{Dimopoulos:2019wew,
archiveprefix = {arXiv},
author = {Dimopoulos, Konstantinos and Markkanen, Tommi and Racioppi, Antonio and Vaskonen, Ville},
doi = {10.1088/1475-7516/2019/07/046},
eprint = {1903.09598},
journal = {JCAP},
pages = {046},
primaryclass = {astro-ph.CO},
reportnumber = {IMPERIAL/TP/2019/TM/01, KCL-PH-TH/2019-33},
title = {{Primordial Black Holes from Thermal Inflation}},
volume = {07},
year = {2019}}


\end{document}